\documentclass[12pt,draftclsnofoot,onecolumn]{IEEEtran}
\usepackage{graphicx}
\usepackage{subfigure}
\usepackage{amsmath,amsthm,amsfonts}
\usepackage{cite}
\usepackage{bm}
\usepackage{url}
\usepackage{array}
\usepackage[colorlinks,linkcolor=black,anchorcolor=blue,citecolor=black]{hyperref}
\usepackage{enumerate}
\usepackage{mdwlist}
\usepackage{amsthm}
\usepackage[T1]{fontenc} % optional T1 font encoding

\allowdisplaybreaks[4]
\theoremstyle{remark}

\begin{document}

\title{Coherent Detection for Short-Packet Physical-Layer Network Coding with FSK Modulation}
\author{Zhaorui~Wang,~\IEEEmembership{Student Member,~IEEE,}
	and~Soung~Chang~Liew,~\IEEEmembership{Fellow,~IEEE}
\thanks{Z. Wang and S. C. Liew are with the Department of Information Engineering, The Chinese University of Hong Kong, Shatin, New Territories, Hong Kong. Email:~\{zrwang2009,~soung\}@ie.cuhk.edu.hk.} 
}
\maketitle
\markboth{Technical Report}{}%

\begin{abstract}
This paper investigates coherent detection for physical-layer network coding (PNC) with short packet transmissions in a two-way relay channel (TWRC). PNC turns superimposed EM waves into network-coded messages to improve throughput in a relay system. To achieve this, accurate channel information at the relay is a necessity. Much prior work applies preambles to estimate the channel. For \emph{long packets}, the preamble overhead is low because of the large data payload. For \emph{short packets}, that is not the case. To avoid excessive overhead, we consider a set-up in which short packets do not have preambles. A key challenge is how the relay can estimate the channel and detect the network-coded messages jointly based on the received signals from the two end users. We design a coherent detector that makes use of a belief propagation (BP) algorithm to do so. For concreteness, we focus on frequency-shift-keying (FSK) modulation. We show how the BP algorithm can be simplified and made practical with Gaussian-mixture passing. In addition, we demonstrate that prior knowledge on the channel distribution is not needed with our framework. Benchmarked against the detector with prior knowledge of the channel distribution, numerical results show that our detector can have nearly the same performance without such prior knowledge.
\end{abstract}

\begin{IEEEkeywords}
Physical-layer network coding, short packet, message passing, frequency shift keying, M2M communications.
\end{IEEEkeywords}

\section{Introduction}
\begin{figure}[ht]
  \centering
  \includegraphics[scale=0.80]{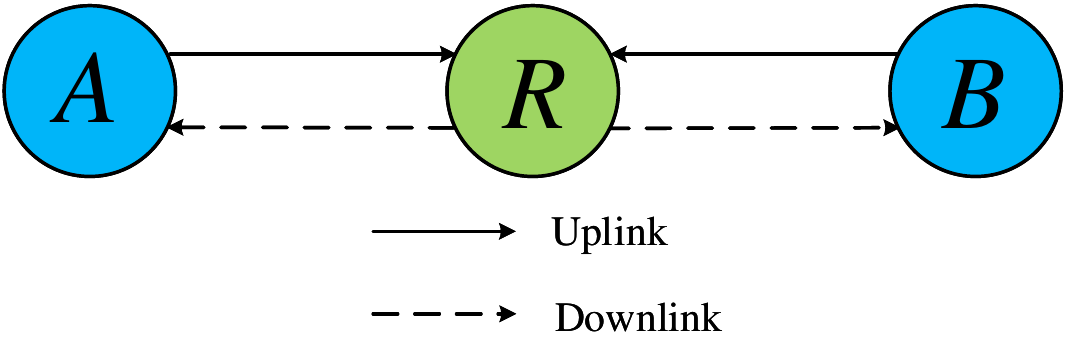}\\
  \caption{A two-way relay channel operated with physical-layer network coding (PNC), where two users \emph{A} and \emph{B} exchange messages via relay \emph{R}.}\label{Fig1}
\end{figure}

This paper investigates coherent detection for physical layer network coding (PNC)\cite{zhang2006hot}\cite{popovski2007physical}\cite{liew2013physical}\cite{nazer2011compute} with short packet transmissions\cite{popovski2018wireless}\cite{popovski2014ultra}\cite{durisi2016toward}\cite{gu2018ultra}\cite{wang2014cellular}. We study a two-way relay channel (TWRC) operated with PNC, as shown in Fig. \ref{Fig1}. Users \emph{A} and \emph{B} are out of each other’s transmission range, and they exchange messages with the assistance of relay \emph{R}. In the uplink phase, users \emph{A} and \emph{B} transmit their messages simultaneously to relay \emph{R}. From the overlapped signals, relay \emph{R} deduces a network-coded message. In the downlink phase, relay \emph{R} broadcasts the network-coded message to both users. User \emph{A} then uses the network-coded message and its own message to deduce the message from user \emph{B}. Likewise for user \emph{B}. We focus on the uplink of PNC. 

This work assumes the use of frequency-shift-keying (FSK) modulation in the PNC system (FSK-PNC). FSK modulated signals are constant envelope signals which can be nonlinearly amplified. In this case, the power amplifier could operate with low back-off and thus achieves high power amplifier efficiency\cite{goldsmith2005wireless}\cite{liang1999nonlinear}. FSK-PNC with short packets could be applied to M2M communications, in which the machines generate tiny messages with stringent energy constraints\cite{durisi2016toward}\cite{lien2011toward}\cite{mukherjee2018energy}. In this scenario, users \emph{A} and \emph{B} are machines exchanging messages through a relay \emph{R}.

PNC turns superimposed EM waves into network-coded messages, greatly improving the throughput in a relay system. Accurate channel information at the relay is a must. Much prior work used preambles to estimate the channel\cite{lu2013implementation}\cite{you2017reliable}\cite{tan2018mobile}\cite{pan2017practical}\cite{pan2018network}\cite{ullah2017phase}. For \emph{long packets}, the preamble overhead is low because of the large data payload. For \emph{short packets}, that is not the case. For example, the packet preamble of IEEE 802.11a has 320 symbols --- a short packet of only several hundred symbols of data payload will incur significant overhead. In this paper, for concreteness we focus on a short packet with 128 symbols. In this case, the overhead is 2.5 times larger than the data payload. To avoid excessive overhead, we consider a set-up in which short packets do not have preambles. A key challenge is how the relay can estimate the channel and detect the network-coded messages jointly based on the received signals from the two end users.

We design a coherent detector that makes use of a belief propagation (BP) algorithm\cite{kschischang2001factor}\cite{loeliger2004introduction}\cite{loeliger2007factor}\cite{zhu2009message} for joint channel estimation and network-coded message detection in FSK-modulated PNC. We show how the BP algorithm can be simplified and made practical with Gaussian-mixture message passing. In particular, we show that the messages propagated in the BP algorithm are Gaussian mixtures. The BP algorithm only needs to pass the means, covariance matrices, and coefficients of the different Gaussian components in the Gaussian mixture, greatly reducing the complexity. To further reduce complexity, we study three different methods to cut down the number of Gaussian components in the mixture. 

This paper first studies a detector that requires prior knowledge on the channel distribution (i.e., the detector has the \emph{a priori} probability distribution of the channels). Benchmarked against an ideal detector that knows the channels exactly (i.e., a detector that does not need to do channel estimation and only has to do detection), our detector only suffers 0.7 dB BER performance loss under Rayleigh \emph{a priori} channel distribution.

This paper next studies a detector that does not require prior knowledge on the channel distribution. The new framework is more versatile and more robust in that it can work under different possible channel distributions. We show that the new detector has nearly the same BER performance as the detector with prior channel distribution knowledge. Thus, the versatility and robustness of the new framework can be obtained without trading off performance.

The remainder of this paper is organized as follows: Section II overviews related works. Section III introduces the system model adopted by this paper. Section IV presents our detector design with prior knowledge on the channel distribution. Section V extends the design to one without the prior knowledge. Numerical results are given in the Subsection of IV-D and the second part of Section V. Section VI concludes this paper. The notations of this paper are summarized in Appendix A.

\section{Related Work}
There have been prior investigations on FSK-PNC detection. Specifically, \cite{yu2016physical} studied coherent detection and \cite{wang2018noncoherent}\cite{wang2018optimal}\cite{sorensen2009physical} \cite{valenti2011noncoherent} studied noncoherent detection. A coherent detector has access to the magnitude and the phase of the received signals, whereas a noncoherent detector only has access to the magnitude of the received signals. Since a coherent detector has all the signal information that a noncoherent detector has, but not vice versa, the coherent detector potentially has better performance.  In this paper, we focus on coherent FSK-PNC detection.

The authors of \cite{yu2016physical} put forth a coherent FSK-PNC detector for power-balanced channels, assuming that the channels are perfectly known at the relay. That is, the detector does not need to do channel estimation and only has to do detection. Although channel estimation can be performed using a preamble in the packet, a preamble would have occupied a large portion of a short packet, causing large overhead. To avoid the overhead, we consider short packets without preambles. In addition, we do not assume power-balanced channel and our detector can work with arbitrary channel distributions.

\section{System Model}
In this paper, for concreteness and as a reference, we assume the bandwidth of our communication system is 1 MHz. Furthermore, we assume the packet length is 128 bits, and thus the packet duration is 128 $\mu s$. We aim for a system employing inexpensive devices. The local oscillators (LOs) in such devices are assumed to be low-cost and thus the frequencies generated by the LO may not be highly accurate. In addition, the RFs at users \emph{A}, \emph{B} and the relay \emph{R} are not synchronized to a common clock. In general, the phase offset of the RFs between user $u \in \left\{ {A,B} \right\}$ and \emph{R} can be expressed as  
\begin{align}
\theta _u^{{\rm{RF}}}\left( t \right) = 2\pi \int\limits_0^t {f_u^{{\rm{RF}}}\left( \tau  \right)d\tau } {\rm{ + }}\varphi _u^{{\rm{RF}}} + \varepsilon _u^{{\rm{RF}}}(t) \label{eq:Sys-1}
\end{align}
where $f_u^{{\rm{RF}}}\left( t \right)$ is the \emph{carrier frequency offset} (CFO) of the RFs between user $u \in \left\{ {A,B} \right\}$ and \emph{R}; $\varphi _u^{{\rm{RF}}}$ is an initial phase offset (the phase offset at the beginning of a packet) between user \emph{u} and \emph{R}; and $\varepsilon _u^{{\rm{RF}}}(t)$ is a random phase offset diffusion due to phase noise.

The CFO $f_u^{{\rm{RF}}}\left( t \right)$ may vary from time to time due to the instability and inaccuracy of the frequency-generating oscillators at user \emph{u} and \emph{R}. However, for short packets of our interest here, the CFO remains more or less constant within the packet duration of 128 $\mu s$ so that we can write
\begin{align}
f_u^{{\rm{RF}}}\left( t \right) = f_u^{{\rm{RF}}}
\end{align}
for a particular packet\cite{tan2018mobile}\cite{wang2018dcap}. In general, the CFO $f_A^{{\rm{RF}}}$ at user \emph{A} is not equal to the CFO $f_B^{{\rm{RF}}}$ at user \emph{B}. Furthermore, the additional phase offset due to random phase noise may not have accumulated during the short packet duration so that we can assume
\begin{align}
\varepsilon _u^{{\rm{RF}}}(t) = 0
\end{align}
for a particular short packet. In short, we assume that the coherence times of the RFs at user \emph{u} and \emph{R} are much larger than 128 $\mu s$. Thus, for a particular packet, \eqref{eq:Sys-1} can be rewritten as 
\begin{align}
\theta _u^{{\rm{RF}}}\left( t \right) = 2\pi f_u^{{\rm{RF}}}t{\rm{ + }}\varphi _u^{{\rm{RF}}}. \label{eq:Sys-2}
\end{align}
\begin{figure*}[]
	\centering
	\includegraphics[scale=1.3]{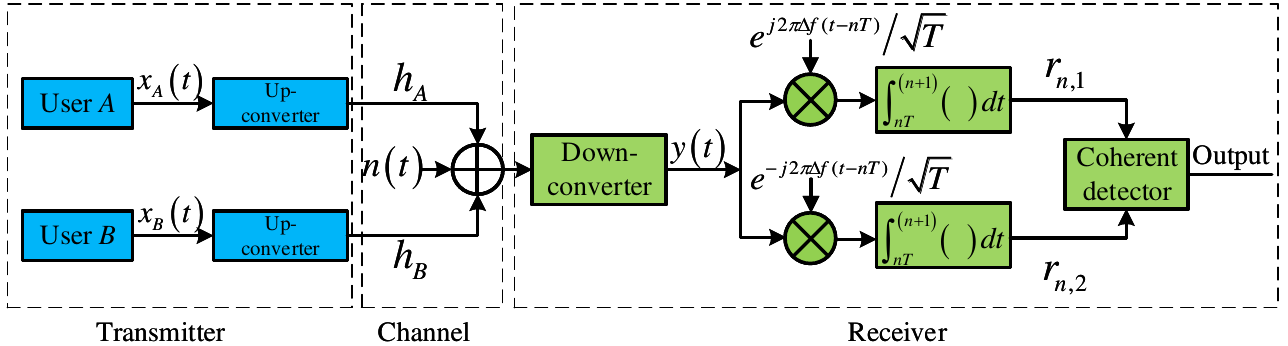}\\
	\caption{Structure of uplink transmissions and reception for a two-way relay channel (TWRC).}\label{Fig2}
\end{figure*}

Fig. \ref{Fig2} shows the structure of uplink transmissions and reception for a two-way relay channel (TWRC). The overall structure of the TWRC coherent receiver is the same as that of a conventional single-user coherent receiver (see \cite{goldsmith2005wireless}\cite{ProakisJohnG2008Dc}) except for the coherent detector at the far right in Fig. \ref{Fig2}. In this paper, we consider a coherent detector that jointly estimates the channel parameters and detects the bit-wise XOR of the messages from users \emph{A} and \emph{B}. The channel parameters being estimated include the channel gains, the phase offsets, and the CFOs associated with users \emph{A} and \emph{B}. In the following, we overview the various processes in Fig. \ref{Fig2}.

\subsection{Baseband Modulator}
This paper assumes both users \emph{A} and \emph{B} adopt continuous phase FSK\cite{goldsmith2005wireless}. FSK encodes bit information into the transmitted frequencies. Specifically, in \emph{M}-ary FSK, there are \emph{M} frequencies to transmit on, and hence \emph{M} possible FSK modulated symbols. Each FSK modulated symbol encodes ${\log _2}M$ bits. For PNC with binary FSK investigated in this paper, users \emph{A} and \emph{B} use the same two frequencies to encode their bits 0 and 1: if users \emph{A} and \emph{B} both transmit bit 0 or both transmit bit 1, then their frequencies overlap; otherwise, their frequencies are distinct.

Let ${s_{n,u}} \in \{ 0,1\}$, $n = 0, \cdots ,N - 1$, be user \emph{u}'s information source bits, where \emph{N} is the packet length. Within each symbol duration $nT \le t < (n + 1)T$, the continuous-phase FSK modulated baseband signal of user \emph{u} can be expressed as
\begin{align}
{x_u}\left( t \right) = 
\begin{cases}
\frac{{\rm{1}}}{{\sqrt T }}{e^{ - j2\pi \Delta f(t - nT) + j\varphi _{n,u}^{\text{CFSK}}}}& \text{if ${s_{n,u}} = 0$}\\
\frac{{\rm{1}}}{{\sqrt T }}{e^{j2\pi \Delta f(t - nT) + j\varphi _{n,u}^{\text{CFSK}}}}& \text{if ${s_{n,u}} = 1$}
\end{cases} \label{eq:Sys-A1} 
\end{align}
where $\frac{{\rm{1}}}{{\sqrt T }}$ is a power normalization factor. The baseband signal uses frequency ${\rm{ - }}\Delta f$ to represent bit 0 and frequency $\Delta f$  to represent bit 1. In this paper, we assume $\Delta f = \frac{1}{{2T}}$\cite{goldsmith2005wireless}. The term $\varphi _{n,u}^{{\rm{CFSK}}}$  corresponds to the phase accumulated over the past $n$  symbol periods given that continuous-phase FSK is used. As such,
\begin{align}
\varphi _{0,u}^{{\rm{CFSK}}} = 0;
\end{align}
\begin{align}
\varphi _{n,u}^{{\rm{CFSK}}} = 2\pi \Delta fT\sum\limits_{i = 0}^{n - 1} {(2{s_{i,u}} - 1)}
\end{align}
for $n \ge 1$\cite{goldsmith2005wireless}\cite{ProakisJohnG2008Dc}. The binary FSK modulated signals  ${x_u}\left( t \right)$ for $u \in \left\{ {A, B} \right\}$ are upconverted to the passband as shown in Fig. \ref{Fig2} before being transmitted. 

\subsection{Baseband Receiver}
In the TWRC uplink, users \emph{A} and \emph{B} transmit their messages to relay \emph{R} simultaneously. Assuming the signal arrival times are aligned, the superimposed signals from users \emph{A} and \emph{B} are down-converted to the baseband as shown in Fig. \ref{Fig2}. The received complex-baseband signal at the receiver is
\begin{align}
y\left( t \right) = \sum\limits_{u \in \left\{ {A, B} \right\}} {{h_u}{{\tilde x}_u}\left( t \right)}  + n\left( t \right) \label{eq:Sys-B1}
\end{align}
where
\begin{align}
{\tilde x_u}\left( t \right) = 
\begin{cases}
\frac{{\rm{1}}}{{\sqrt T }}{e^{j2\pi \left( { - \Delta f + f_u^{{\rm{RF}}}} \right)\left( {t - nT} \right) + j{\theta _{n,u}}}}& \text{if ${s_{n,u}} = 0$}\\
\frac{{\rm{1}}}{{\sqrt T }}{e^{j2\pi \left( {\Delta f + f_u^{{\rm{RF}}}} \right)(t - nT) + j{\theta _{n,u}}}}& \text{if ${s_{n,u}} = 1$}
\end{cases}. \label{eq:Sys-B2}
\end{align}
In \eqref{eq:Sys-B1}, ${h_u}$ is the channel between user \emph{u} and relay \emph{R}. In this paper, we assume the channel to be flat-slow-fading: specifically ${h_u}$  remains constant within one packet duration. In addition, the channels ${h_A}$ and ${h_B}$ are independent. The noise $n(t)$ is white Gaussian noise with zero mean and double-sided \emph{power spectral density} (PSD) $N_0/2$. In \eqref{eq:Sys-B2}, due to oscillator asynchrony, the phase $2n\pi f_u^{{\rm{RF}}}T$  caused by CFO and initial phase offset $\varphi _u^{{\rm{RF}}}$  are added to ${\tilde x_u}\left( t \right)$  within the phase term
\begin{align}
{\theta _{n,u}} = \varphi _{n,u}^{{\rm{CFSK}}}{\rm{ + }}2n\pi f_u^{{\rm{RF}}}T + \varphi _u^{{\rm{RF}}}.
\end{align} 

The received signal $y\left( t \right)$  is cross-correlated with the two reference frequencies in each symbol period as shown in Fig. \ref{Fig2}. In the symbol period $nT \le t < (n + 1)T$, $n = 0, \cdots ,N - 1$, the outputs of the first and second branches are
\begin{align}
\begin{array}{l}
{r_{n,1}} = \frac{1}{{\sqrt T }}\int\limits_{nT}^{(n + 1)T} {{e^{j2\pi \Delta f(t - nT)}}y\left( t \right)} dt\\
{r_{n,2}} = \frac{1}{{\sqrt T }}\int\limits_{nT}^{(n + 1)T} {{e^{ - j2\pi \Delta f(t - nT)}}y\left( t \right)} dt
\end{array}. \label{eq:Sys-B3}
\end{align}

Let ${{\bf{r}}_n} = {\left( {{r_{n,1}},{r_{n,2}}} \right)^{\rm{T}}}$, $n = 0, \cdots ,N - 1$, where ${\left[  \bullet  \right]^{\rm{T}}}$ is a transpose operator. The signal vector can be expressed as

\begin{align}
\begin{array}{l}
{{\bf{r}}_n} = \\
\begin{cases}
{\left( {{h_A}{\alpha _A}{e^{j{\theta _{n,A}}}} + {h_B}{\alpha _B}{{\mathop{\rm e}\nolimits} ^{j{\theta _{n,B}}}} + {w_{n,1}},{h_A}{\beta _A}{e^{j{\theta _{n,A}}}} + {h_B}{\beta _B}{{\mathop{\rm e}\nolimits} ^{j{\theta _{n,B}}}} + {w_{n,2}}} \right)^{\rm{T}}}& \text{if ${s_{n,A}}= 0,{s_{n,B}} = 0$}\\
{\left( {{h_A}{\beta _A}{e^{j{\theta _{n,A}}}} + {h_B}{\beta _B}{{\mathop{\rm e}\nolimits} ^{j{\theta _{n,B}}}} + {w_{n,1}},{h_A}{\alpha _A}{e^{j{\theta _{n,A}}}} + {h_B}{\alpha _B}{{\mathop{\rm e}\nolimits} ^{j{\theta _{n,B}}}} + {w_{n,2}}} \right)^{\rm{T}}}& \text{if ${s_{n,A}}= 1,{s_{n,B}} = 1$}\\
{\left( {{h_A}{\alpha _A}{e^{j{\theta _{n,A}}}} + {h_B}{\beta _B}{{\mathop{\rm e}\nolimits} ^{j{\theta _{n,B}}}} + {w_{n,1}},{h_A}{\beta _A}{e^{j{\theta _{n,A}}}} + {\kern 1pt} {h_B}{\alpha _B}{{\mathop{\rm e}\nolimits} ^{j{\theta _{n,B}}}} + {w_{n,2}}} \right)^{\rm{T}}}& \text{if ${s_{n,A}}= 0,{s_{n,B}} = 1$}\\
{\left( {{\kern 1pt} {h_A}{\beta _A}{e^{j{\theta _{n,A}}}} + {h_B}{\alpha _B}{{\mathop{\rm e}\nolimits} ^{j{\theta _{n,B}}}} + {w_{n,1}},{h_A}{\alpha _A}{e^{j{\theta _{n,A}}}} + {h_B}{\beta _B}{{\mathop{\rm e}\nolimits} ^{j{\theta _{n,B}}}} + {w_{n,2}}} \right)^{\rm{T}}}& \text{if ${s_{n,A}}= 1,{s_{n,B}} = 0$}\\
\end{cases}
\end{array} \label{eq:Sys-B4}
\end{align}
where ${\alpha _u} = \frac{{{e^{j2\pi f_u^{{\rm{RF}}}T}} - 1}}{{j2\pi f_u^{{\rm{RF}}}T}}$, ${\beta _u} = \frac{{{e^{j2\pi f_u^{{\rm{RF}}}T}} - 1}}{{{\rm{ - }}j2\pi  + j2\pi f_u^{{\rm{RF}}}T}}$, $u \in \left\{ {A,B} \right\}$; ${w_{n,1}}$ and ${w_{n,2}}$ are complex Gaussian random variables with mean zero and covariance $N_0$.  

In practice, we should have a rough idea of the range of the CFOs $f_A^{{\rm{RF}}}$  and $f_B^{{\rm{RF}}}$. This can be acquired through equipment testing beforehand\cite{tan2018mobile}\cite{wang2018dcap}. The range of CFO investigated in this paper is $f_u^{{\rm{RF}}} \in \left[ { - 10{\kern 1pt} {\rm{Hz}},10{\kern 1pt} {\rm{kHz}}} \right]$ for $u \in \left\{ {A,B} \right\}$ (the oscillators in software-defined radio boards, for example,  have CFO smaller than this range\cite{tan2018mobile}\cite{wang2018dcap}). In addition, the bandwidth is 1 MHz, and thus symbol duration $T{\rm{ = }}1 {\kern 1pt} us$. In this case, we have ${\alpha _A} \approx 1$, ${\alpha _B} \approx 1$, ${\beta _A} \approx 0$, and ${\beta _B} \approx 0$ in \eqref{eq:Sys-B4}. In addition, let ${h_{n,u}} = {h_u}{e^{j{\theta _{n,u}}}}$ for $u \in \left\{ {A,B} \right\}$, 
${{\bf{h}}_n} = {\left[ 
{\begin{array}{*{20}{c}}
{{h_{n,A}}}&{{h_{n,B}}}
\end{array}} \right]^{\rm{T}}}$,
and 
${{\bf{w}}_n} = {\left[
{\begin{array}{*{20}{c}}
{{w_{n,1}}}&{{w_{n,2}}}
\end{array}} \right]^{\rm{T}}}$.
Then, \eqref{eq:Sys-B4} becomes
\begin{align}
{{\bf{r}}_n} = {{\bf{Z}}_{\left( {{s_{n,A}},{s_{n,B}}} \right)}}{{\bf{h}}_n}{\rm{ + }}{{\bf{w}}_n} \label{eq:Sys-B4.1}
\end{align}
where ${{\bf{w}}_n}$ is a complex Gaussian random noise vector with zero mean and covariance matrix 
${\sum _{\bf{w}}} = \left[ {\begin{array}{*{20}{c}}
	{{N_0}}&0\\
	0&{{N_0}}
\end{array}} \right]$, and ${{\bf{Z}}_{\left( {{s_{n,A}},{s_{n,B}}} \right)}}$ is a matrix fully determined by the transmitted bits $\left( {{s_{n,A}},{s_{n,B}}} \right)$. In particular, when $\left( {{s_{n,A}},{s_{n,B}}} \right){\rm{ = }}\left( {{\rm{0,0}}} \right)$, ${{\bf{Z}}_{\left( {0,0} \right)}} = \left[ {\begin{array}{*{20}{c}}
1&1\\
0&0
\end{array}} \right]$; when $\left( {{s_{n,A}},{s_{n,B}}} \right){\rm{ = }}\left( {{\rm{1,1}}} \right)$, ${{\bf{Z}}_{\left( {1,1} \right)}} = \left[ {\begin{array}{*{20}{c}}
0&0\\
1&1
\end{array}} \right]$; when $\left( {{s_{n,A}},{s_{n,B}}} \right){\rm{ = }}\left( {{\rm{0,1}}} \right)$, ${{\bf{Z}}_{\left( {0,1} \right)}} = \left[ {\begin{array}{*{20}{c}}
1&0\\
0&1
\end{array}} \right]$; when $\left( {{s_{n,A}},{s_{n,B}}} \right){\rm{ = }}\left( {{\rm{1,0}}} \right)$, ${{\bf{Z}}_{\left( {1,0} \right)}} = \left[ {\begin{array}{*{20}{c}}
0&1\\
1&0
\end{array}} \right]$. 

In \eqref{eq:Sys-B4.1}, ${h_{n,u}}$ changes  with $n$. The ratio between   ${h_{n,u}}$ and ${h_{n + {\rm{1}},u}}$  is
\begin{align}
\frac{{{h_{n + 1,u}}}}{{{h_{n,u}}}} = {e^{j\left( {{\theta _{n + 1,u}} - {\theta _{n,u}}} \right)}} = {e^{j\left( {{\partial _1}{{\tilde s}_{n,u}} + {\partial _2}f_u^{{\rm{RF}}}} \right)}} \label{eq:Sys-D5}
\end{align}
where constants ${\partial _1} =  - 2\pi \Delta fT$ and ${\partial _2} = 2\pi T$, and ${\tilde s_{n,u}} = {\rm{1}} - 2{s_{n,u}}$. Let 
\begin{align}
{{\bf{G}}_n} = {\rm{diag}}\left( {{e^{j\left( {{\partial _1}{{\tilde s}_{n,A}} + {\partial _2}f_A^{{\rm{RF}}}} \right)}},{e^{j\left( {{\partial _1}{{\tilde s}_{n,B}} + {\partial _2}f_B^{{\rm{RF}}}} \right)}}} \right),
\end{align}
the relationship between ${{\bf{h}}_{n - 1}}$ and ${{\bf{h}}_n}$ can be expressed as 
\begin{align}
	{{\bf{h}}_n} = {{\bf{G}}_{n - 1}}{{\bf{h}}_{n - 1}}. \label{eq:Sys-D7}
\end{align}
Note that, the statistic of ${{\bf{h}}_n}$ now is determined not only by the statistics of ${h_A}$ and ${h_B}$ but also by the statistics of the initial phase offsets, the CFOs, and the transmitted bits from \emph{A} and \emph{B}. In addition, in the rest of this paper, by “channel” we mean ${{\bf{h}}_n}$, and by ``pure channel'', we mean ${\left[ {\begin{array}{*{20}{c}}
		{{h_A}}&{{h_B}}
		\end{array}} \right]^{\rm{T}}}$.

\subsection{Coherent Detector}
The detector applies PNC coherent detection that detects the bit-wise XOR of the message of \emph{A} and the message of \emph{B}, ${s_{n,R}} = {s_{n,A}} \oplus {s_{n,B}}$, $n = 0, \cdots ,N - 1$, based on the received signals $\left( {{{\bf{r}}_0}, \cdots ,{{\bf{r}}_{N - 1}}} \right)$. In particular, the coherent detector makes decisions based not only on the magnitudes, but also on the phases of the received signals. We assume each user independently transmits bits 0 and 1 with equal probability so that ${\rm{Pr}}\left( {{s_{n,R}} = 0} \right) = {\rm{Pr}}\left( {{s_{n,R}} = 1} \right) = 1/2$, where $\Pr \left(  \bullet  \right)$ is the \emph{probability mass function} (PMF) of discrete random variables. We abbreviate the vector $\left( {{{\bf{r}}_i},{{\bf{r}}_{i + 1}}, \cdots ,{{\bf{r}}_{j - 1}},{{\bf{r}}_j}} \right)$ by the notation ${{\bf{r}}_{i:j}}$. In this paper, similar abbreviations also apply to other vector notations. Let ${\rm{Pr}}\left( {\left. {{s_{n,R}}} \right|{{\bf{r}}_{0:N - 1}}} \right)$ be the conditional PMF of the \emph{n}-th XORed symbol given the signals of the overall received packet. The coherent detector detects the XORed symbol based on the \emph{maximum a posteriori probability} (MAP) criterion:
\begin{equation}
\begin{split}
s_{n,R}^ * &= \mathop {{\rm{arg max }}}\limits_{x \in \left\{ {0,1} \right\}} {\rm{Pr}}\left( {\left. {{s_{n,R}} = x} \right|{{\bf{r}}_{0:N - 1}}} \right)\\
&= \mathop {{\rm{arg max }}}\limits_{x \in \left\{ {0,1} \right\}} \sum\limits_{({s_{n,A}},{s_{n,B}}):{s_{n,R}} = x} {{\rm{Pr}}\left( {\left. {{s_{n,A}},{s_{n,B}}} \right|{{\bf{r}}_{0:N - 1}}} \right)} 
\end{split} \label{eq:Sys-C1}
\end{equation}
where $s_{n,R}^ * $ is the decision on the XORed symbol ${s_{n,R}}$ of users \emph{A} and \emph{B}. The second line shows that, to compute ${\rm{Pr}}\left( {\left. {{s_{n,R}} = x} \right|{{\bf{r}}_{0:N - 1}}} \right)$, we need to sum ${\rm{Pr}}\left( {\left. {{s_{n,A}},{s_{n,B}}} \right|{{\bf{r}}_{0:N - 1}}} \right)$  over $({s_{n,A}},{s_{n,B}})$  such that ${s_{n,R}}{\rm{ = }}{s_{n,A}} \oplus {s_{n,B}} = x$.  

The analysis in the rest of this paper is based on the system model described by \eqref{eq:Sys-B4.1} and \eqref{eq:Sys-D7}, and the detector in \eqref{eq:Sys-C1}. Note that, after the transformation of the channels shown above, the detector does not need to estimate the initial phase offsets of users \emph{A} and \emph{B} separately from their channels since the initial phase offsets are incorporated into the channels. 

We need to jointly detect the transmitted signals, and estimate channels and CFOs of users \emph{A} and \emph{B}. The next section presents a detector that computes the PMF   in \eqref{eq:Sys-C1} assuming the \emph{a priori} distributions of channel parameters are available. Section V then presents a detector that computes ${\rm{Pr}}\left( {\left. {{s_{n,A}},{s_{n,B}}} \right|{{\bf{r}}_{0:N - 1}}} \right)$  without such \emph{a priori} distributions.

\section{Detector Design with Prior Distributions of Channel Parameters} 
In much prior work on PNC detection, the channels were either assumed to be perfectly known\cite{zhang2006hot}\cite{popovski2007physical}\cite{liew2013physical}\cite{shi2016subtleties}\cite{shi2017complex} or estimated through preambles and pilots\cite{lu2013implementation}\cite{you2017reliable}\cite{tan2018mobile}\cite{pan2017practical}\cite{pan2018network}\cite{ullah2017phase}. This paper assumes a set-up without preambles. Based on the received data symbols only, joint channel estimation, CFO estimation, and data detection are performed. As indicated in \eqref{eq:Sys-D7}, channels at different time instances, ${\left\{ {{{\bf{h}}_n}} \right\}_{n = 0:N - 1}}$, are related deterministically through ${\left\{ {{{\bf{G}}_n}} \right\}_{n = 0:N - {\rm{2}}}}$. In addition, we assume CFOs are constant within the whole short packet. As such, the received signals over the whole data payload contain common information about the channels and the CFOs. In particular, the information related to the channels and the CFOs extracted from the other symbols are useful for detection of a particular symbol ${s_{n,R}}$. 

We write the PMF in \eqref{eq:Sys-C1} as
\begin{equation}
\begin{split}
&{\rm{Pr}}\left( {\left. {{{\bf{s}}_n}} \right|{{\bf{r}}_{0:N - 1}}} \right)\\
&= \int {d{\bf{f}}d{{\bf{h}}_n}} p\left( {\left. {{{\bf{s}}_n},{{\bf{h}}_n},{\bf{f}}} \right|{{\bf{r}}_{0:N - 1}}} \right)\\
&= \int {d{\bf{f}}} p\left( {\left. {{{\bf{s}}_n},{\bf{f}}} \right|{{\bf{r}}_{0:N - 1}}} \right)
\end{split} \label{eq:IV-1}
\end{equation}
where ${{\bf{s}}_n} = \left( {{s_{n,A}},{s_{n,B}}} \right)$ and ${\bf{f}} = \left( {f_A^{{\rm{RF}}},f_B^{{\rm{RF}}}} \right)$. The integration range of CFO is $f_u^{{\rm{RF}}} \in \left[ { - 10{\kern 1pt} {\rm{k}}{\rm{Hz}},10{\kern 1pt} {\rm{kHz}}} \right]$, $u \in \left\{ {A,B} \right\}$. The integration range of ${{\bf{h}}_n}$, on the other hand, is the whole complex plane. A closed form of the integral in \eqref{eq:IV-1} is not readily available. However, for a fixed ${\bf{f}}$, a closed form of the integral over ${{\bf{h}}_n}$  can be obtained. Thus, our strategy is to first approximate the following integration 
\begin{align}
{\rm{Pr}}\left( {\left. {{{\bf{s}}_n}} \right|{{\bf{r}}_{0:N - 1}}} \right) = \int {d{\bf{f}}} p\left( {\left. {{{\bf{s}}_n},{\bf{f}}} \right|{{\bf{r}}_{0:N - 1}}} \right) \label{eq:IV-6.1}
\end{align}
with the summation
\begin{align}
{\rm{Pr}}\left( {\left. {{{\bf{s}}_n}} \right|{{\bf{r}}_{0:N - 1}}} \right) \approx \kappa \sum\limits_i {p\left( {\left. {{{\bf{s}}_n},{{\bf{f}}_i}} \right|{{\bf{r}}_{0:N - 1}}} \right)}  \label{eq:IV-7}
\end{align}
where $\kappa $  is the quantized step in the summation and ${{\bf{f}}_i}$  is the \emph{i}-th quantized value of ${\bf{f}}$.

Later in this section, we will show that $p\left( {\left. {{{\bf{s}}_n},{{\bf{h}}_n},{{\bf{f}}_i}} \right|{{\bf{r}}_{0:N - 1}}} \right)$  for a given fixed ${{\bf{f}}_i}$  is a mixture (sum) of Gaussian functions in ${{\bf{h}}_n}$. A message passing algorithm can then be constructed based on the passing of the means and covariance matrices of the Gaussian functions in ${{\bf{h}}_n}$, allowing $p\left( {\left. {{{\bf{s}}_n},{{\bf{h}}_n},{{\bf{f}}_i}} \right|{{\bf{r}}_{0:N - 1}}} \right)$ to be integrated in closed form to yield $p\left( {\left. {{{\bf{s}}_n},{{\bf{f}}_i}} \right|{{\bf{r}}_{0:N - 1}}} \right)$  in \eqref{eq:IV-7}.

\subsection{Computation through Belief Propagation Algorithm}
This subsection shows $p\left( {\left. {{{\bf{s}}_n},{{\bf{h}}_n},{{\bf{f}}_i}} \right|{{\bf{r}}_{0:N - 1}}} \right)$ (and hence also $p\left( {\left. {{{\bf{s}}_n},{{\bf{f}}_i}} \right|{{\bf{r}}_{0:N - 1}}} \right)$) can be computed by a Belief Propagation (BP) algorithm. We first show the decomposition of $p\left( {\left. {{{\bf{s}}_n},{{\bf{h}}_n},{{\bf{f}}_i}} \right|{{\bf{r}}_{0:N - 1}}} \right)$ that allows a BP message-passing algorithm to be constructed. In the following, we drop the index \emph{i} in ${{\bf{f}}_i}$, with the understanding that the BP algorithm will be run over ${{\bf{f}}_i}$  for different \emph{i}  and then \eqref{eq:IV-7} will be applied in the final step to tally $p\left( {\left. {{{\bf{s}}_n},{{\bf{f}}_i}} \right|{{\bf{r}}_{0:N - 1}}} \right)$  over different \emph{i} to obtain ${\rm{Pr}}\left( {\left. {{{\bf{s}}_n}} \right|{{\bf{r}}_{0:N - 1}}} \right)$. 

\begin{figure}[ht]
	\centering
	\includegraphics[scale=1.00]{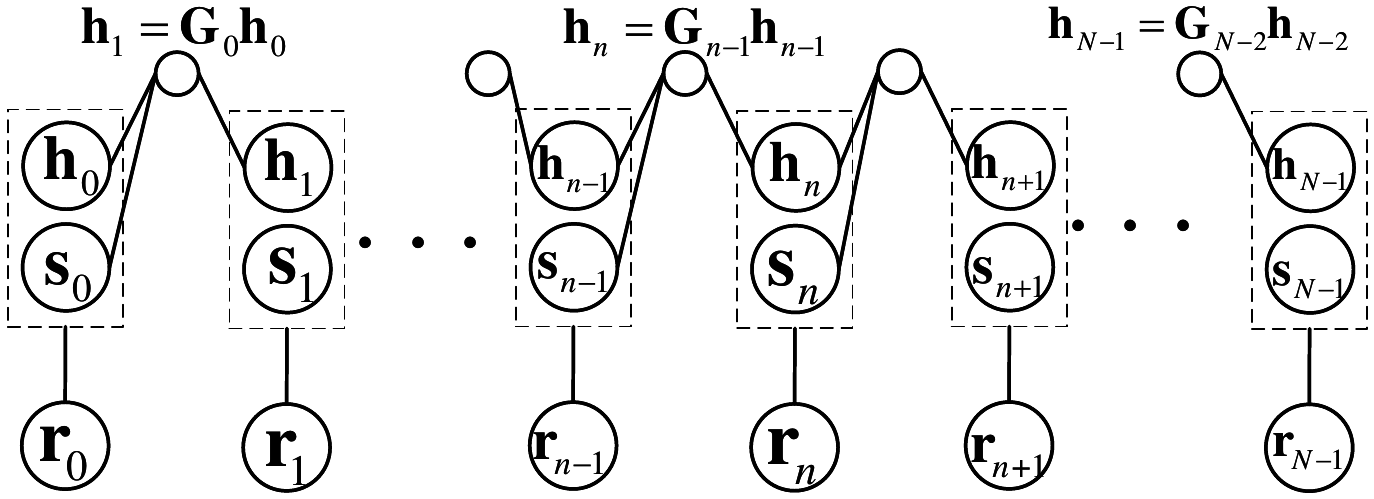}\\
	\caption{Graphical interpretation of $p\left( {\left. {{{\bf{s}}_n},{{\bf{h}}_n},{\bf{f}}} \right|{{\bf{r}}_{0:N - 1}}} \right)$  in \eqref{eq:IV-2} for a fixed $\bf{f}$. Note that ${{\bf{G}}_i}$  is a function of ${{\bf{s}}_i}$  and $\bf{f}$. }\label{Fig6}
\end{figure}

Appendix B shows that the integrand in \eqref{eq:IV-1} can be expressed as
\begin{equation}
\begin{split}
&p\left( {\left. {{{\bf{s}}_n},{{\bf{h}}_n},{\bf{f}}} \right|{{\bf{r}}_{0:N - 1}}} \right)\\
&\propto p({{\bf{h}}_n},{\bf{f}})\underbrace {\prod\limits_{i = {\rm{0}}}^{n - 1} {\sum\limits_{{\kern 1pt} {\kern 1pt} {\kern 1pt} {\kern 1pt} {\kern 1pt} {{\bf{s}}_i}} {p({{\bf{r}}_i}|{{\bf{h}}_i},{{\bf{s}}_i}){\psi _{{{\bf{G}}_i}}}({{\bf{h}}_{i + 1}},{{\bf{h}}_i})} } }_{{\rm M}_r^{\left( n \right)}\left( {{{\bf{h}}_n}} \right)}p({{\bf{r}}_n}|{{\bf{s}}_n},{{\bf{h}}_n})\underbrace {\prod\limits_{i = n{\rm{ + 1}}}^{N - {\rm{1}}} {\sum\limits_{{\kern 1pt} {\kern 1pt} {\kern 1pt} {\kern 1pt} {\kern 1pt} {{\bf{s}}_i}} {{\psi _{{{\bf{G}}_{i - 1}}}}({{\bf{h}}_i},{{\bf{h}}_{i - 1}})p({{\bf{r}}_i}|{{\bf{h}}_i},{{\bf{s}}_i})} } }_{{\rm M}_l^{\left( n \right)}\left( {{{\bf{s}}_n},{{\bf{h}}_n}} \right)}
\end{split} \label{eq:IV-2}
\end{equation}
where  $p({{\bf{h}}_n},\bf{f})$ is the prior distribution of the channel parameters, and ${\psi _G}(x,y)$  is an indicator function such that ${\psi _G}(x,y) = 1$  if  $x = Gy$ and ${\psi _G}(x,y) = 0$  otherwise. Fig. \ref{Fig6} shows a graphical interpretation of \eqref{eq:IV-2}, from which we can obtain a BP algorithm for the computation of $p\left( {\left. {{{\bf{s}}_n},{{\bf{h}}_n},{\bf{f}}} \right|{{\bf{r}}_{0:N - 1}}} \right)$. 

In Fig. \ref{Fig6}, ${{\bf{r}}_i}$, $i = 0, \cdots ,N - 1$ are the observations; $\left( {{{\bf{s}}_i},{{\bf{h}}_i}} \right)$, $i = 0, \cdots ,N - 1$ are the hidden variables to be detected and estimated. The hidden variables are not independent and are related through the function 
\begin{align}
{{\bf{h}}_{i + 1}} = {{\bf{G}}_i}{{\bf{h}}_i} \label{eq:IV-2.1}
\end{align}
where the matrix  ${{\bf{G}}_i}$, as described in Section III-B, is a function of ${{\bf{s}}_i}$  and ${\bf{f}}$. With respect to the BP algorithm, ${\rm M}_r^{\left( n \right)}\left( {{{\bf{h}}_n}} \right)$  (${\rm M}_l^{\left( n \right)}\left( {{{\bf{s}}_n},{{\bf{h}}_n}} \right)$) in \eqref{eq:IV-2} can be interpreted as a right-bound (left-bound) message coming from the left (right), incident on node \emph{n}. At node \emph{n}, besides the messages from other nodes,  ${\rm M}_r^{\left( n \right)}\left( {{{\bf{h}}_n}} \right)$ and ${\rm M}_l^{\left( n \right)}\left( {{{\bf{s}}_n},{{\bf{h}}_n}} \right)$, we also have messages  $p\left( {\left. {{{\bf{r}}_n}} \right|{{\bf{s}}_n},{{\bf{h}}_n}} \right)$  and $p({{\bf{h}}_n},\bf{f})$  generated at node \emph{n}. All these messages are used together for the overall computation of $p\left( {\left. {{{\bf{s}}_n},{{\bf{h}}_n},{\bf{f}}} \right|{{\bf{r}}_{0:N - 1}}} \right)$  as expressed in \eqref{eq:IV-2}. 

\subsection{Representation of Right/Left-bound Message by Gaussian Mixture}
The essence of the BP algorithm is the iterative computation of ${\rm M}_r^{\left( n \right)}\left( {{{\bf{h}}_n}} \right)$  and ${\rm M}_l^{\left( n \right)}\left( {{{\bf{s}}_n},{{\bf{h}}_n}} \right)$ for successive \emph{n}. To compute $p\left( {\left. {{{\bf{s}}_n},{\bf{f}}} \right|{{\bf{r}}_{0:N - 1}}} \right)$  in \eqref{eq:IV-7} through the integration of $p\left( {\left. {{{\bf{s}}_n},{{\bf{h}}_n},{\bf{f}}} \right|{{\bf{r}}_{0:N - 1}}} \right)$  over ${{\bf{h}}_n}$  for a fixed ${{\bf{s}}_n}$  and ${\bf{f}}$, potentially we need to execute the BP algorithm for every realization of ${{\bf{h}}_n}$. This results in large computation complexity since ${{\bf{h}}_n}$  is continuous random variable over the whole complex plane. 

Fortunately, we could solve the problem by leveraging the fact that the left/right-bound messages are Gaussian mixtures in ${{\bf{h}}_n}$, and that the means and covariance matrices associated with the Gaussian components fully characterize ${\rm M}_r^{\left( n \right)}\left( {{{\bf{h}}_n}} \right)$ and ${\rm M}_l^{\left( n \right)}\left( {{{\bf{s}}_n},{{\bf{h}}_n}} \right)$ as continuous functions in ${{\bf{h}}_n}$. We explain the recursive computation of ${\rm M}_r^{\left( n \right)}\left( {{{\bf{h}}_n}} \right)$ and ${\rm M}_l^{\left( n \right)}\left( {{{\bf{s}}_n},{{\bf{h}}_n}} \right)$ below, and along the way, we show why they are Gaussian mixtures. 

Let us focus on ${\rm M}_l^{\left( n \right)}\left( {{{\bf{s}}_n},{{\bf{h}}_n}} \right)$ (the treatment for ${\rm M}_r^{\left( n \right)}\left( {{{\bf{h}}_n}} \right)$ is similar). ${\rm M}_l^{\left( n \right)}\left( {{{\bf{s}}_n},{{\bf{h}}_n}} \right)$ in \eqref{eq:IV-2} can be expressed as a recursion of message passing from right to left:
\begin{align}
{\rm M}_l^{\left( n \right)}({{\bf{s}}_n},{{\bf{h}}_n}) = \sum\limits_{{\kern 1pt} {\kern 1pt} {\kern 1pt} {\kern 1pt} {\kern 1pt} {{\bf{s}}_{n + 1}}} {{\psi _{{{\bf{G}}_n}}}({{\bf{h}}_{n + 1}},{{\bf{h}}_n})p({{\bf{r}}_{n + 1}}|{{\bf{h}}_{n + 1}},{{\bf{s}}_{n + 1}}){\rm M}_l^{\left( {n{\rm{ + 1}}} \right)}({{\bf{s}}_{n{\rm{ + 1}}}},{{\bf{h}}_{n{\rm{ + 1}}}})} \label{eq:IV-B1}
\end{align}
where ${\rm M}_l^{\left( {n{\rm{ + 1}}} \right)}({{\bf{s}}_{n{\rm{ + 1}}}},{{\bf{h}}_{n{\rm{ + 1}}}})$  is the left-bound message incident on node (\emph{n}+1). At node (\emph{n}+1), to compute ${\rm M}_l^{\left( n \right)}({{\bf{s}}_n},{{\bf{h}}_n})$  for passing to $\left( {{{\bf{s}}_n},{{\bf{h}}_n}} \right)$, we first combine node (\emph{n}+1)'s message $p({{\bf{r}}_{n + 1}}|{{\bf{h}}_{n + 1}},{{\bf{s}}_{n + 1}})$ and the left bound-message ${\rm M}_l^{\left( {n{\rm{ + 1}}} \right)}({{\bf{s}}_{n{\rm{ + 1}}}},{{\bf{h}}_{n{\rm{ + 1}}}})$  from node (\emph{n}+2), and then rotate the channel ${{\bf{h}}_{n + 1}}$  to  ${{\bf{h}}_n}$ through ${\psi _{{{\bf{G}}_n}}}({{\bf{h}}_{n + 1}},{{\bf{h}}_n})$, and finally sum them over different ${{\bf{s}}_{n + 1}}$.

By exchanging the order of summation of multiplication, ${\rm M}_l^{\left( n \right)}\left( {{{\bf{s}}_n},{{\bf{h}}_n}} \right)$ in \eqref{eq:IV-2} can also be written as
\begin{equation}
\begin{split}
&{\rm M}_l^{\left( n \right)}\left( {{{\bf{s}}_n},{{\bf{h}}_n}} \right)\\
&= \sum\limits_{{\kern 1pt} {\kern 1pt} {\kern 1pt} {\kern 1pt} {\kern 1pt} {{\bf{s}}_{n + 1:N - 1}}} {p({{\bf{r}}_{n + 1}}|{{\bf{h}}_{n + 1}},{{\bf{s}}_{n + 1}}) \cdots p({{\bf{r}}_{N - 1}}|{{\bf{h}}_{N - 1}},{{\bf{s}}_{N - 1}}){\psi _{{{\bf{G}}_n}}}({{\bf{h}}_{n + 1}},{{\bf{h}}_n}) \cdots {\psi _{{{\bf{G}}_{N - 2}}}}({{\bf{h}}_{N - 1}},{{\bf{h}}_{N - 2}})} 
\end{split}. \label{eq:IV-B1.1}
\end{equation}
In \eqref{eq:IV-B1.1}, the Gaussian PDFs of ${{\bf{r}}_i}$,
$p({{\bf{r}}_i}|{{\bf{h}}_i},{{\bf{s}}_i})$, $i = n + 1, \cdots ,N - 1$, could be interpreted as Gaussian functions of ${{\bf{h}}_i}$\footnote{A Gaussian PDF has a proper normalized form of $\frac{{\rm{1}}}{{{{\left( {{\rm{2}}\pi } \right)}^{\frac{d}{2}}}{{\left| \sum  \right|}^{\frac{1}{2}}}}}\exp \left\{ { - \frac{1}{2}{{\left( {{\bf{Z}} - {\bf{\mu }}} \right)}^ * }{\sum ^{ - 1}}\left( {{\bf{Z}} - {\bf{\mu }}} \right)} \right\}$, where \emph{d} is the dimension of the Gaussian random variable  ${\bf{Z}}$,  ${\bf{\mu }}$ is the mean of the ${\bf{Z}}$, and $\sum $ is the covariance matrix of ${\bf{Z}}$; a Gaussian function has the same form but may not be normalized to be a proper PDF. For example, $p({{\bf{r}}_i}|{{\bf{h}}_i},{{\bf{s}}_i})$   is a Gaussian PDF of ${{\bf{r}}_i}$,  but a Gaussian function of ${{\bf{h}}_i}$.}. We note that through ${\psi _{{{\bf{G}}_n}}}({{\bf{h}}_{n + 1}},{{\bf{h}}_n}) \cdots {\psi _{{{\bf{G}}_{N - 2}}}}({{\bf{h}}_{N - 1}},{{\bf{h}}_{N - 2}})$, each Gaussian function $p({{\bf{r}}_i}|{{\bf{h}}_i},{{\bf{s}}_i})$ in ${{\bf{h}}_i}$, $i > n$, becomes a Gaussian function in ${{\bf{h}}_n}$  in the end. Alternatively,  $p({{\bf{r}}_i}|{{\bf{h}}_i},{{\bf{s}}_i})$ can be expressed as a Gaussian PDF in  ${{\bf{h}}_n}$ multiplied by a weight (coefficient). Furthermore, the product of Gaussian functions $p({{\bf{r}}_i}|{{\bf{h}}_i},{{\bf{s}}_i})$, $i = n + 1, \cdots ,N - 1$  could be expressed as a Gaussian PDF in ${{\bf{h}}_n}$  multiplied by a coefficient\cite{petersen2008matrix} that depends on the specific values of ${{\bf{s}}_{n + 1:N-1}}$ and  ${\bf{f}}$. Considering all possible values of ${{\bf{s}}_{n + 1:N - 1}}$  in the summation in \eqref{eq:IV-B1.1}, the left-bound message is therefore a mixture of weighted Gaussian PDFs. The PDF of a mixture of Gaussian components are fully determined by the means, covariance matrices, and the corresponding coefficients of the underlying Gaussian components. In this case, instead of computing the whole PDFs of the left/right-bound message through the BP algorithm for each specific realization of  ${{\bf{h}}_n}$, we could just compute the mean, covariance matrix, and the corresponding coefficient of each Gaussian component of the right/left-bound message. The complexity mentioned above is thus greatly reduced. 

Given the analysis above, the rest of this subsection details the computation of the means, the covariance matrixes, and the corresponding coefficients of the Gaussian components. 

We focus on the left-bound message ${\rm M}_l^{\left( n \right)}({{\bf{s}}_n},{{\bf{h}}_n})$. According to the analysis above,  ${\rm M}_l^{\left( n \right)}({{\bf{s}}_n},{{\bf{h}}_n})$ can be expressed as
\begin{align}
{\rm M}_l^{\left( n \right)}({{\bf{s}}_n},{{\bf{h}}_n}) = \sum\limits_i {\rho _i^{(n|n{\rm{ + }}1)}CN\left( {{\bf{h}}_n^{},{\bf{\bar h}}_i^{\left( {n|n + 1} \right)},\Sigma _i^{\left( {n|n{\rm{ + 1}}} \right)}} \right)} \label{eq:IV-C2}
\end{align}
where  $CN\left( {{\bf{h}}_n^{},{\bf{\bar h}}_i^{\left( {n|n + 1} \right)},\Sigma _i^{\left( {n|n{\rm{ + 1}}} \right)}} \right)$ represents a specific Gaussian component of  ${\rm M}_l^{\left( n \right)}({{\bf{s}}_n},{{\bf{h}}_n})$ in ${\bf{h}}_n^{}$ with mean ${\bf{\bar h}}_i^{\left( {n|n + 1} \right)}$  and covariance matrix $\Sigma _i^{\left( {n|n{\rm{ + 1}}} \right)}$ (``\emph{n}'' is the subscript of ${\bf{h}}_n^{}$  and ``(\emph{n}+1)'' is the initial subscript of ${{\bf{r}}_{n + 1:N - 1}}$, the observations on which the left-bound messages depend). In addition,  $\rho _i^{(n|n{\rm{ + }}1)}$ is the coefficient of the Gaussian component. We next show how to compute the mean  ${\bf{\bar h}}_i^{\left( {n|n + 1} \right)}$, covariance matrix  $\Sigma _i^{\left( {n|n{\rm{ + 1}}} \right)}$, and the coefficient $\rho _i^{(n|n{\rm{ + }}1)}$  of ${\rm M}_l^{\left( n \right)}({{\bf{s}}_n},{{\bf{h}}_n})$  in \eqref{eq:IV-C2} from the mean ${\bf{\bar h}}_i^{\left( {n + 1|n + 2} \right)}$, covariance matrix  $\Sigma _i^{\left( {n{\rm{ + }}1|n{\rm{ + }}2} \right)}$, and the coefficient $\rho _i^{(n + 1|n + 2)}$  of ${\rm M}_l^{\left( {n{\rm{ + 1}}} \right)}({{\bf{s}}_{n{\rm{ + 1}}}},{{\bf{h}}_{n{\rm{ + 1}}}})$. To this end, we first write
\begin{align}
{\rm M}_l^{\left( {n{\rm{ + 1}}} \right)}({{\bf{s}}_{n{\rm{ + 1}}}},{{\bf{h}}_{n{\rm{ + 1}}}}) = \sum\limits_i {\rho _i^{(n + 1|n + 2)}} CN\left( {{\bf{h}}_{n + 1}^{},{\bf{\bar h}}_i^{\left( {n + 1|n + 2} \right)},\Sigma _i^{\left( {n{\rm{ + }}1|n{\rm{ + }}2} \right)}} \right). \label{eq:IV-C4.1}
\end{align}
In this case,
\begin{equation}
\begin{split}
&{\rm M}_l^{\left( n \right)}({{\bf{s}}_n},{{\bf{h}}_n})\\
&= \sum\limits_{{\kern 1pt} {\kern 1pt} {\kern 1pt} {\kern 1pt} {\kern 1pt} {{\bf{s}}_{n + 1}}} {{\psi _{{{\bf{G}}_n}}}({{\bf{h}}_{n + 1}},{{\bf{h}}_n})p({{\bf{r}}_{n + 1}}|{{\bf{h}}_{n + 1}},{{\bf{s}}_{n + 1}})} \sum\limits_i {\rho _i^{(n + 1|n + 2)}} CN\left( {{\bf{h}}_{n + 1}^{},{\bf{\bar h}}_i^{\left( {n + 1|n + 2} \right)},\Sigma _i^{\left( {n{\rm{ + }}1|n{\rm{ + }}2} \right)}} \right)\\
&= \sum\limits_{{\kern 1pt} {\kern 1pt} {\kern 1pt} {\kern 1pt} {\kern 1pt} {{\bf{s}}_{n + 1}}} {\sum\limits_i {\underbrace {\underbrace {\rho _i^{(n + 1|n + 2)}p({{\bf{r}}_{n + 1}}|{{\bf{h}}_{n + 1}},{{\bf{s}}_{n + 1}})CN\left( {{\bf{h}}_{n + 1}^{},{\bf{\bar h}}_i^{\left( {n + 1|n + 2} \right)},\Sigma _i^{\left( {n{\rm{ + }}1|n{\rm{ + }}2} \right)}} \right)}_{{\Theta _1}}{\psi _{{{\bf{G}}_n}}}({{\bf{h}}_{n + 1}},{{\bf{h}}_n})}_{{\Theta _2}}} } 
\end{split}. \label{eq:IV-C5}
\end{equation}

From \eqref{eq:IV-C5}, the computation of the mean, covariance matrix, and the coefficient of each Gaussian component of  ${\rm M}_l^{\left( n \right)}({{\bf{s}}_n},{{\bf{h}}_n})$ can be performed through the following two steps:
\begin{enumerate}
	\item \emph{Write the summand in \eqref{eq:IV-C5} as a Gaussian function in ${\bf{h}}_{n{\rm{ + }}1}^{}$}.\\
	Specifically, we express the summand ${\Theta _1}$ in \eqref{eq:IV-C5} as
	\begin{align}
	{\Theta _1}{\rm{ = }}\rho _i^{(n{\rm{ + 1}}|n{\rm{ + }}1)}CN\left( {{\bf{h}}_{n + 1}^{},{\bf{\bar h}}_i^{\left( {n + 1|n + 1} \right)},\Sigma _i^{\left( {n{\rm{ + }}1|n{\rm{ + 1}}} \right)}} \right). \label{eq:IV-C6}
	\end{align}
	To this end, we first write $p\left( {\left. {{{\bf{r}}_{n + 1}}} \right|{{\bf{s}}_{n{\rm{ + 1}}}},{\bf{h}}_{n + 1}^{}} \right)$  in \eqref{eq:IV-C5} as a Gaussian function in  ${\bf{h}}_{n + 1}^{}$ with mean ${{\bf{\bar h}}^{{{\bf{s}}_{n + 1}}}}$, inverse covariance matrix  ${\left( {{\Sigma ^{{{\bf{s}}_{n + 1}}}}} \right)^{ - 1}}$, and a constant ${c^{{{\bf{s}}_{n + 1}}}}$. The details are given in Appendix C. Then, according to\cite{petersen2008matrix},
	\begin{align}
	{\left( {\Sigma _i^{\left( {n + {\rm{1}}|n + 1} \right)}} \right)^{ - 1}} = {\left( {{\Sigma ^{{{\bf{s}}_{n + 1}}}}} \right)^{ - 1}} + {\left( {\Sigma _i^{\left( {n + 1|n + 2} \right)}} \right)^{ - 1}}; \label{eq:IV-C7}
	\end{align}
	\begin{align}
	{\bf{\bar h}}_i^{\left( {n + {\rm{1}}|n + 1} \right)} = \Sigma _i^{\left( {n + {\rm{1}}|n + 1} \right)}\left( {{{\left( {{\Sigma ^{{{\bf{s}}_{n + 1}}}}} \right)}^{ - 1}}{{{\bf{\bar h}}}^{{{\bf{s}}_{n + 1}}}} + {{\left( {\Sigma _i^{\left( {n + 1|n + 2} \right)}} \right)}^{ - 1}}{\bf{\bar h}}_i^{\left( {n + 1|n + 2} \right)}} \right); \label{eq:IV-C8}
	\end{align}
	and
	\begin{align}
     \begin{split}
     &\rho _i^{(n + 1|n + {\rm{1}})} = \frac{{{{\left| {\Sigma _i^{\left( {n + {\rm{1}}|n + 1} \right)}} \right|}^{\frac{1}{2}}}}}{{{{\left| {\Sigma _i^{\left( {n + 1|n + 2} \right)}} \right|}^{\frac{1}{2}}}}}\rho _i^{(n + 1|n + 2)}{c^{{{\bf{s}}_{n + 1}}}}\\
     &\times \exp \left\{ \begin{array}{l}
     \frac{{\rm{1}}}{{\rm{2}}}{\left( {{{\left( {{\Sigma ^{{{\bf{s}}_{n + 1}}}}} \right)}^{ - 1}}{{{\bf{\bar h}}}^{{{\bf{s}}_{n + 1}}}} + {{\left( {\Sigma _i^{\left( {n + 1|n + 2} \right)}} \right)}^{ - 1}}{\bf{\bar h}}_i^{\left( {n + 1|n + 2} \right)}} \right)^ * }{\bf{\bar h}}_i^{\left( {n + {\rm{1}}|n + 1} \right)}\\
     - \frac{{\rm{1}}}{{\rm{2}}}\left( {{{\left( {{{{\bf{\bar h}}}^{{{\bf{s}}_{n + 1}}}}} \right)}^ * }{{\left( {{\Sigma ^{{{\bf{s}}_{n + 1}}}}} \right)}^{ - 1}}{{{\bf{\bar h}}}^{({{\bf{s}}_{n + 1}})}} + {{\left( {{\bf{\bar h}}_i^{\left( {n + 1|n + 2} \right)}} \right)}^ * }{{\left( {\Sigma _i^{\left( {n + 1|n + 2} \right)}} \right)}^{ - 1}}{\bf{\bar h}}_i^{\left( {n + 1|n + 2} \right)}} \right)
     \end{array} \right\}
     \end{split}. \label{eq:IV-C9}
   	\end{align}
	\item \emph{Write the summand in \eqref{eq:IV-C5} as a Gaussian function in ${\bf{h}}_{n}^{}$}.\\
	Specifically, we express the summand ${\Theta _2}$ in \eqref{eq:IV-C5} as
	\begin{align}
	 {\Theta _2} = \rho _i^{(n|n + {\rm{1}})}CN\left( {{\bf{h}}_n^{},{\bf{\bar h}}_i^{\left( {n|n + 1} \right)},\Sigma _i^{\left( {n|n{\rm{ + 1}}} \right)}} \right). \label{eq:IV-C10}
	\end{align}
	The updated mean, covariance matrix, and the coefficient are as follows:
	\begin{align}
	 {\bf{\bar h}}_i^{\left( {n|n{\rm{ + }}1} \right)}{\rm{ = }}{\bf{G}}_n^{{\rm{ - 1}}}{\bf{\bar h}}_i^{\left( {n{\rm{ + 1}}|n{\rm{ + }}1} \right)}; \label{eq:IV-C11}
	\end{align}
	\begin{align}
	 {\left( {\Sigma _i^{\left( {n|n{\rm{ + }}1} \right)}} \right)^{ - 1}} = {\left( {{\bf{G}}_n^{}} \right)^ * }{\left( {\Sigma _i^{\left( {n{\rm{ + 1}}|n{\rm{ + }}1} \right)}} \right)^{ - 1}}{\bf{G}}_n^{}; \label{eq:IV-C12}
	\end{align}
	and
	\begin{align}
	\rho _i^{(n|n + {\rm{1}})} = \rho _i^{(n + 1|n + {\rm{1}})}.\label{eq:IV-C13}
	\end{align}
	Note that for numerical stability, $\rho _i^{(n|n + {\rm{1}})}$  should be normalized such that 
	\begin{align}
		\sum\limits_i {\rho _i^{(n|n{\rm{ + }}1)}} {\rm{ = 1}} \label{eq:C13.5}
	\end{align}
	before passing of the message.
\end{enumerate}
After the above procedure, we then get
\begin{align}
\begin{split}
&{\rm M}_l^{\left( n \right)}({{\bf{s}}_n},{{\bf{h}}_n})\\
&= \sum\limits_{{{\bf{s}}_{n{\rm{ + 1}}}}} {\sum\limits_i {\rho _i^{(n|n{\rm{ + }}1)}CN\left( {{\bf{h}}_n^{},{\bf{\bar h}}_i^{\left( {n|n + 1} \right)},\Sigma _i^{\left( {n|n{\rm{ + 1}}} \right)}} \right)} } \\
&= \sum\limits_j {\rho _j^{(n|n{\rm{ + }}1)}CN\left( {{\bf{h}}_n^{},{\bf{\bar h}}_j^{\left( {n|n + 1} \right)},\Sigma _j^{\left( {n|n{\rm{ + 1}}} \right)}} \right)} 
\end{split}. \label{eq:IV-C14}
\end{align}
In this case, the number of terms under the summation is increased by 4 times.

Similarly, we could also compute the mean, covariance matrix, and the coefficient of each Gaussian component of
\begin{align}
{\rm M}_r^{\left( n \right)}\left( {{{\bf{h}}_n}} \right) \propto \sum\limits_i {\rho _i^{(n|n - {\rm{1}})}CN\left( {{\bf{h}}_n^{},{\bf{\bar h}}_i^{\left( {n|n - 1} \right)},\Sigma _i^{\left( {n|n - {\rm{1}}} \right)}} \right)}. \label{eq:IV-C15}
\end{align}

The above procedure describes the computation of the mean, covariance matrix, and the coefficient of each Gaussian component of the left-bound message and the right-bound message. However, from \eqref{eq:IV-C14} and \eqref{eq:IV-C15}, the number of Gaussian components within the right/left-bound messages increase exponentially as the messages are passed from node to node, inducing huge computational complexity. To solve the problem, in each iteration, we curtail the number of the Gaussian components and use the Gaussian mixture residual (\emph{GMR}) components (i.e., the remaining components after the curtailment) to approximate the right/left-bound messages. In this case, the complexity of the detector is in the order of ${\rm{4}} \times GMR \times N$. 

After running the BP algorithm to obtain $p\left( {\left. {{{\bf{s}}_n},{{\bf{h}}_n},{\bf{f}}} \right|{{\bf{r}}_{0:N - 1}}} \right)$ as expressed in \eqref{eq:IV-2}, we then integrate it over ${{\bf{h}}_n}$ to get 
\begin{align}
p\left( {\left. {{{\bf{s}}_n},{\bf{f}}} \right|{{\bf{r}}_{0:N - 1}}} \right) = \int {d{{\bf{h}}_n}} p\left( {\left. {{{\bf{s}}_n},{{\bf{h}}_n},{\bf{f}}} \right|{{\bf{r}}_{0:N - 1}}} \right). \label{eq:IV-6}
\end{align}
In particular, with respect to \eqref{eq:IV-2}, within  $p\left( {\left. {{{\bf{s}}_n},{{\bf{h}}_n},{\bf{f}}} \right|{{\bf{r}}_{0:N - 1}}} \right)$, the message $p({{\bf{r}}_n}|{{\bf{s}}_n},{{\bf{h}}_n})$  can also be expressed as a Gaussian function of ${{\bf{h}}_n}$. The product of  $p({{\bf{r}}_n}|{{\bf{s}}_n},{{\bf{h}}_n})$, the left-bound message, and the right-bound message is still a mixture of Gaussian components. If the prior information  $p({{\bf{h}}_n},\bf{f})$ is available, we have to multiply the Gaussian mixture by the prior information $p({{\bf{h}}_n},\bf{f})$  to get   $p\left( {\left. {{{\bf{s}}_n},{{\bf{h}}_n},{\bf{f}}} \right|{{\bf{r}}_{0:N - 1}}} \right)$ before the integration. If  $p({{\bf{h}}_n},\bf{f})$ is further a Gaussian function in ${{\bf{h}}_n}$, then  $p\left( {\left. {{{\bf{s}}_n},{{\bf{h}}_n},{\bf{f}}} \right|{{\bf{r}}_{0:N - 1}}} \right)$ is still a Gaussian mixture. Note that the integration of each Gaussian function, which can be expressed as a coefficient times a Gaussian PDF, is given by the coefficient. We will detail the detector design in Section V if the prior information $p({{\bf{h}}_n},\bf{f})$ is not available.

For the integration in \eqref{eq:IV-6}, we assume the CFOs are uniformly distributed within a range (see Subsection III-B). In addition, we assume that the pure channels ${h_A}$  and ${h_B}$ are Rayleigh distributed. In particular,  ${h_u}$ for $u \in \left\{ {A,B} \right\}$  is complex Gaussian random variable with zero mean and variance $\sigma _{{h_u}}^2$. Since after phase rotation on the channel in \eqref{eq:Sys-D7}, the distribution of the channels is unchanged,  ${{\bf{h}}_n}$, $n = 0, \cdots ,N - 1$, are still Rayleigh distributed with zero mean and covariance matrix ${\sum _{\bf{h}}} = \left[ {\begin{array}{*{20}{c}}
	{\sigma _{{h_A}}^2}&0\\
	0&{\sigma _{{h_B}}^2}
	\end{array}} \right]$. In this case,
\begin{align}
p({{\bf{h}}_n},{\bf{f}}) \propto p({{\bf{h}}_n}).\label{eq:IV-C15.1}
\end{align}
Substituting \eqref{eq:IV-C14}, \eqref{eq:IV-C15}, and \eqref{eq:IV-C15.1} into \eqref{eq:IV-2} and then \eqref{eq:IV-6}, we have
\begin{align}
\begin{split}
&p\left( {\left. {{{\bf{s}}_n},{\bf{f}}} \right|{{\bf{r}}_{0:N - 1}}} \right)\\
&= \int {d{{\bf{h}}_n}} p\left( {\left. {{{\bf{s}}_n},{{\bf{h}}_n},{\bf{f}}} \right|{{\bf{r}}_{0:N - 1}}} \right)\\
&\approx \sum\limits_{i,j} {\rho _i^{(n|n - {\rm{1}})}\rho _j^{(n|n{\rm{ + }}1)}}\\ &~~~~\times\int {p\left( {{{\bf{h}}_n}} \right)CN\left( {{\bf{h}}_n^{},{\bf{\bar h}}_i^{\left( {n|n - 1} \right)},\Sigma _i^{\left( {n|n - {\rm{1}}} \right)}} \right)CN\left( {{\bf{h}}_n^{},{\bf{\bar h}}_j^{\left( {n|n + 1} \right)},\Sigma _j^{\left( {n|n{\rm{ + 1}}} \right)}} \right)p\left( {\left. {{{\bf{r}}_n}} \right|{{\bf{s}}_n},{{\bf{h}}_n}} \right)d{{\bf{h}}_n}} 
\end{split}. \label{eq:IV-C16}
\end{align}
Note that, the second line is not strictly equal to the third line since we curtail some Gaussian components in \eqref{eq:IV-C14} and \eqref{eq:IV-C15}. We refer to our designed detector as \textbf{``Brief Propagation Coherent Detector (BPCD)''}. In addition, we refer to the Gaussian component reduction technique in this subsection as Curtailment.

\subsection{Reduction of Components through (i) Gaussian Approximation and (ii) Hybrid of Curtailment and Gaussian Approximation}
This subsection presents two additional methods to reduce the number of Gaussian components within the right/left-bound messages in \eqref{eq:IV-C14} and \eqref{eq:IV-C15}.

\emph{\textbf{Gaussian Approximation:}} This method approximates the Gaussian mixture as a single Gaussian PDF instead of curtailing the number of the Gaussian components within the mixture. In this case, the left-bound message (the treatment for the right-bound message is similar) becomes
\begin{align}
\begin{split}
&{\rm M}_l^{\left( n \right)}({{\bf{s}}_n},{{\bf{h}}_n})\\
&= \sum\limits_j {\rho _j^{(n|n{\rm{ + }}1)}CN\left( {{\bf{h}}_n^{},{\bf{\bar h}}_j^{\left( {n|n + 1} \right)},\Sigma _j^{\left( {n|n{\rm{ + 1}}} \right)}} \right)} \\
&\approx CN\left( {{\bf{h}}_n^{},{\bf{\bar h}}_{{\rm{app}}}^{\left( {n|n + 1} \right)},\Sigma _{{\rm{app}}}^{\left( {n|n{\rm{ + 1}}} \right)}} \right) 
\end{split} \label{eq:IV-C1}
\end{align}
where
\begin{align}
{\bf{\bar h}}_{{\rm{app}}}^{\left( {n|n + 1} \right)}{\rm{ = }}\sum\limits_j {\rho _j^{(n|n{\rm{ + }}1)}{\bf{\bar h}}_j^{\left( {n|n + 1} \right)}} \label{eq:C2.1}
\end{align}
and
\begin{align}
\Sigma _{{\rm{app}}}^{\left( {n|n{\rm{ + 1}}} \right)}{\rm{ = }}\sum\limits_j {\rho _j^{(n|n{\rm{ + }}1)}\left[ {\Sigma _j^{\left( {n|n{\rm{ + 1}}} \right)}{\rm{ + }}\left( {{\bf{\bar h}}_j^{\left( {n|n + 1} \right)} - {\bf{\bar h}}_{{\rm{app}}}^{\left( {n|n + 1} \right)}} \right){{\left( {{\bf{\bar h}}_j^{\left( {n|n + 1} \right)} - {\bf{\bar h}}_{{\rm{app}}}^{\left( {n|n + 1} \right)}} \right)}^ * }} \right]}. \label{eq:IV-C3} 
\end{align}
The complexity of BPCD by Gaussian Approximation is in the order of ${\rm{4}}N$.

\emph{\textbf{Hybrid of Curtailment and Gaussian Approximation:}} This method first keeps the $GMR-1$ Gaussian components with the largest coefficients. It then approximates the remaining Gaussian components as a Gaussian function. Thus, altogether we still have \emph{GMR} Gaussian components. The complexity of BPCD Hybrid is in the order of ${\rm{4}} \times GMR \times N$.

\subsection{Numerical Results}
This subsection presents the simulation results of the three different variants of BPCD. BPCD in this section requires \emph{a priori} distributions of the channels and CFOs. In particular, we assume in our simulation that the channel between user \emph{A} and relay \emph{R}, and the channel between user \emph{B} and relay \emph{R}, are both Rayleigh-fading channels with $E\left( {{{\left| {{h_A}} \right|}^2}} \right) = E\left( {{{\left| {{h_B}} \right|}^2}} \right) = 1$. In addition, we assume that CFOs are uniformly distributed. The packet size is 128 bits and the symbol duration is $T{\rm{ = }}1{\kern 1pt} us$. We first study BPCD in which the number of Gaussian components is reduced by Curtailment. After that, we compare the performance of BPCD with different component reduction methods.

\textbf{\emph{1) BER of the Curtailment Method with Different GMRs.}}

We first study the performance of BPCD approximation by Curtailment. In particular, we are interested in the influence of \emph{GMR} on BER under different SNRs in BPCD. We benchmark BPCD against a detector with perfect knowledge of the two users' CFOs, channels, and phases (i.e., the initial phases, the phases accumulated due to CFSK modulation, and the phases due to CFOs). We refer to this detector as \textbf{``perfect coherent detector (PerfCD)''}.
\begin{figure}[ht]
	\centering
	\includegraphics[scale=0.5]{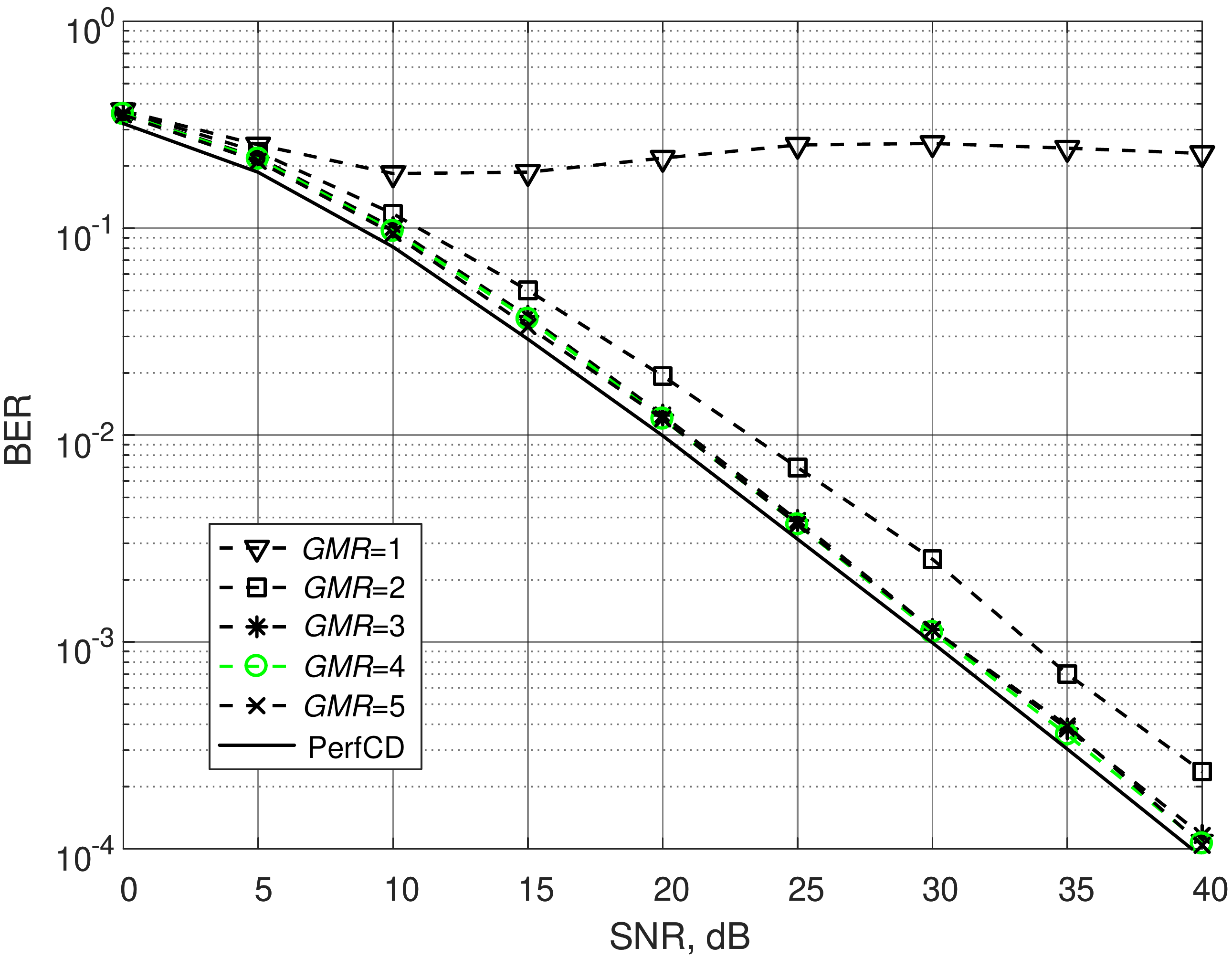}\\
	\caption{BER of PerfCD and BPCD approximation by Curtailment with different \emph{GMR}s}\label{Fig7}
\end{figure}

The results are shown in Fig. \ref{Fig7}. From Fig. \ref{Fig7}, the BER of BPCD with \emph{GMR}=1 is around 0.5. It indicates that keeping only one Gaussian component within the left/right-bound messages is not sufficient. BPCD with \emph{GMR}=2 improves drastically. BPCD with \emph{GMR}=3 has around 2.9 dB SNR gain over that with \emph{GMR}=2 at ${\rm{BER = 1}}{{\rm{0}}^{{\rm{ - 3}}}}$. The performance gaps among BPCD with \emph{GMR}=3, BPCD with \emph{GMR}=4, and BPCD with \emph{GMR}=5 are near zero. Moreover, benchmarked against PerfCD, BPCD with \emph{GMR}=4 (or that with \emph{GMR}=3 or \emph{GMR}=5) has a 0.7 dB performance gap\footnote{The performance gap is due to the estimation error of ${{\bf{h}}_n}$  and $\bf{f}$. We also study the performance of BPCD with a packet size of 512 bits. Numerical results show that the performance gap between BPCD and PerfCD can be further narrowed in that case. We omit the results here since we only focus on the short packet with packet size of 128 bits.}. From the observations and the analysis above, we conclude that BPCD with \emph{GMR}=4 is sufficient for good performance.

\textbf{\emph{2) Comparison of Three Component-Reduction Methods.}} 

\begin{figure}[ht]
	\centering
	\subfigure[BER of PerfCD, BPCD approximation by Curtailment with \emph{GMR}=4, BPCD approximation by Gaussian approximation, and BPCD approximation by Hybrid of Curtailment and Gaussian approximation with \emph{GMR}=4.]
	{
		\label{Fig8a}
		\includegraphics[width=0.6\columnwidth]{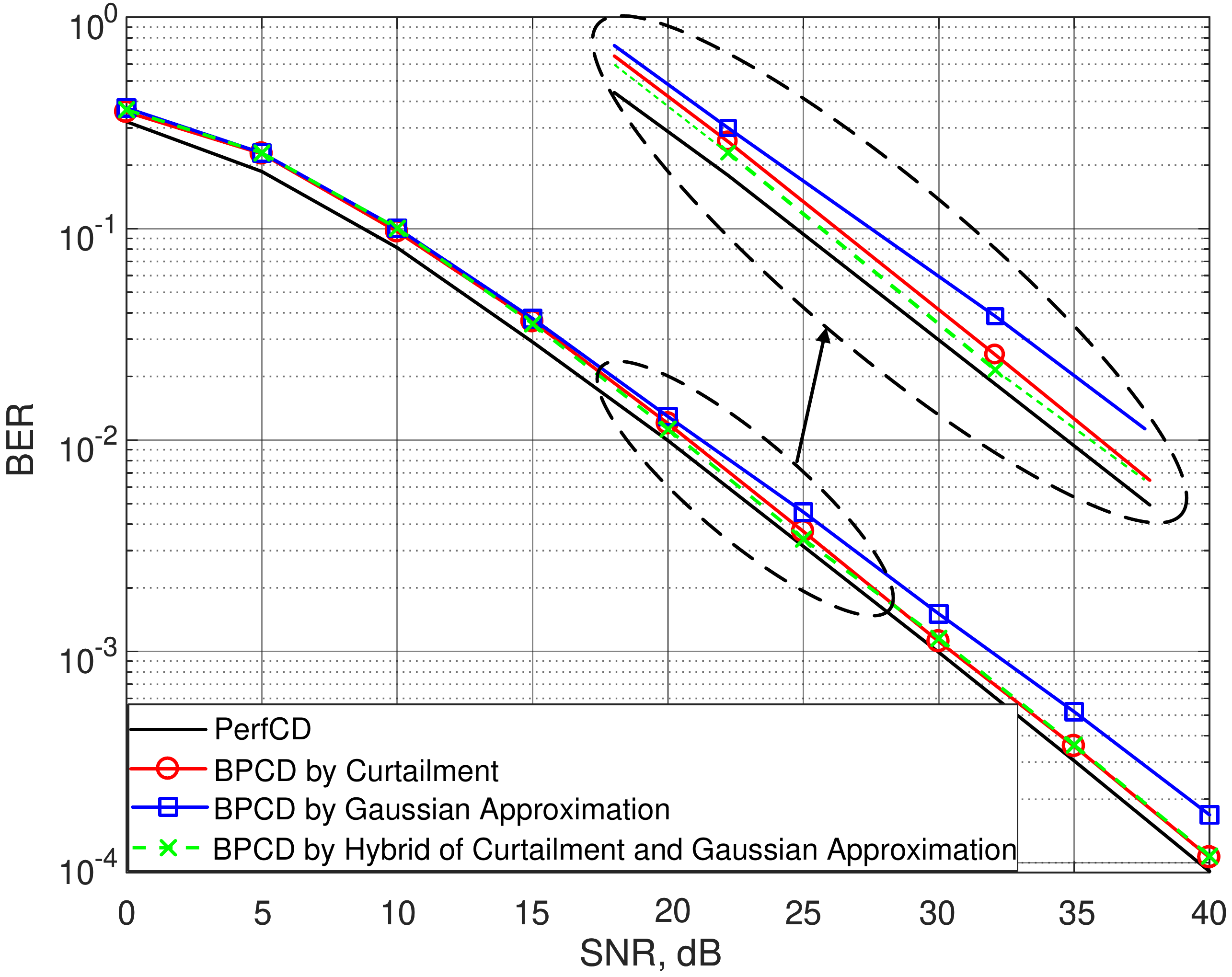}
	}
	\subfigure[MSE of the channel ${{\bf{h}}_n}$.]
	{
		\label{Fig8b}
		\includegraphics[width=0.6\columnwidth]{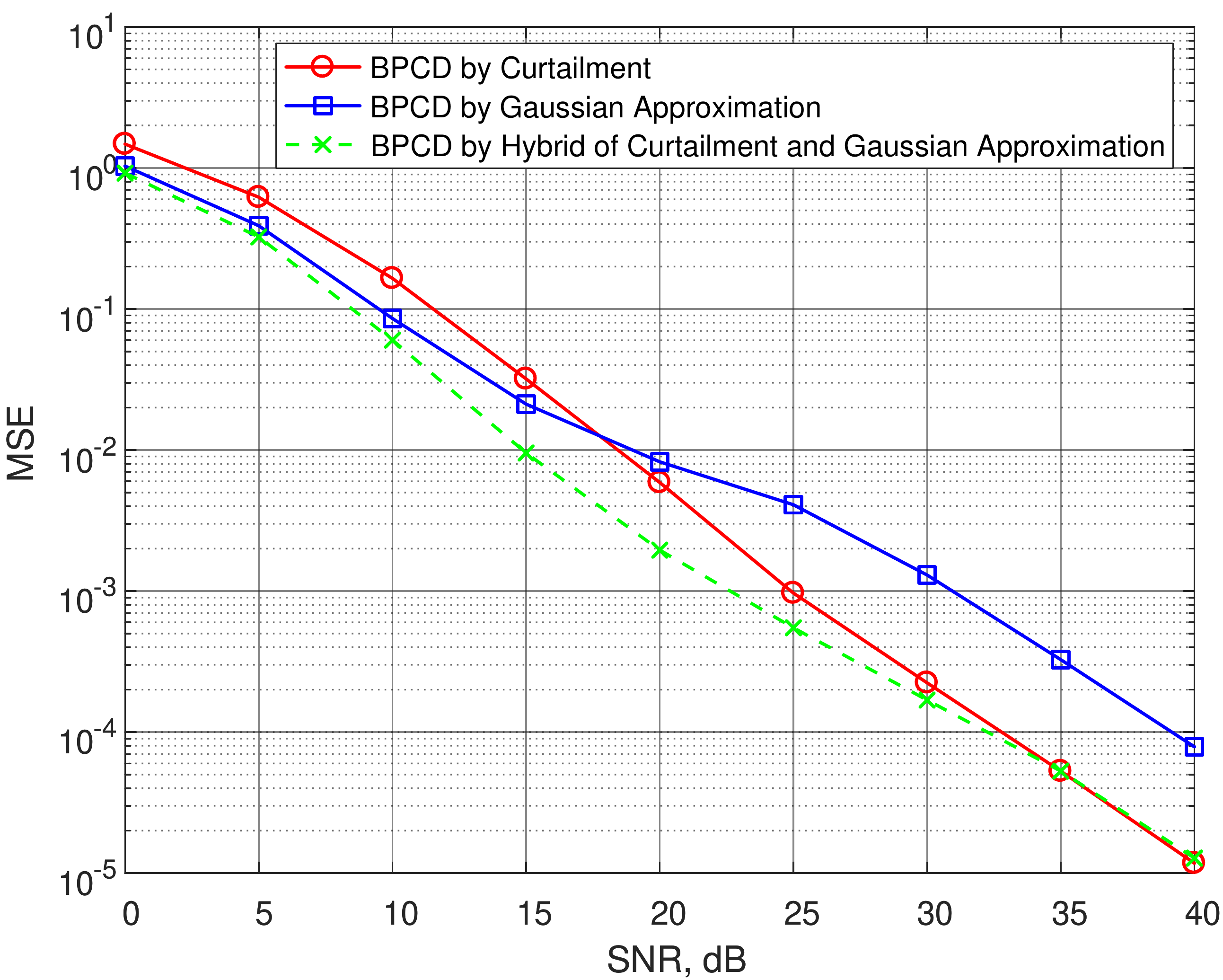}
	}
	\caption{BER and MSE of different variants of BPCD.}\label{Fig8}
\end{figure}

We next study the BER performance of the three component-reduction methods under different SNRs. The results are shown in Fig. \ref{Fig8a}. It can be seen from Fig. \ref{Fig8a} that in the low SNR regime, BPCD by Gaussian Approximation performs as well as BPCD approximation by Curtailment (with \emph{GMR}=4). In the middle and high SNR regime, BPCD by Curtailment outperforms BPCD by Gaussian Approximation. Moreover, BPCD by Hybrid of Curtailment and Gaussian Approximation (with \emph{GMR}=4) performs as well as BPCD approximation by Curtailment in the low and high SNR regime, and BPCD Hybrid performs the best in the middle SNR regime.

To explain the observations in Fig. \ref{Fig8a}, we compute the mean square error (MSE) of the estimations of channels and CFOs. We focus on channel ${{\bf{h}}_n}$  for the \emph{n}-th symbol. The MSE is the same for all \emph{n} since the MSE for each \emph{n} is computed based on the same received packet. The MSE is defined as
\begin{align}
{\rm{MSE = \mathbb{E}}}\left( {{{\left\| {{{\bf{h}}_n} - {\bf{h}}_n^ * } \right\|}^2}} \right)  \label{eq:IV-D1}
\end{align}
where ${\bf{h}}_n^ * $  is the estimation of the channel ${{\bf{h}}_n}$. From Fig. \ref{Fig8b}, BPCD by Gaussian Approximation outperforms BPCD by Curtailment by around 2 dB in the low SNR regime, while BPCD through Curtailment outperforms BPCD through Gaussian Approximation by around 6 dB in the middle and high SNR regime. In the low SNR regime, the variance of each Gaussian component is large and the relative significance of different Gaussian components does not vary much. In this case, Gaussian approximation may be a better way to capture the characteristics of the Gaussian components. The MSE results in Fig. \ref{Fig8b} also indicates that BPCD Hybrid performs the best in the low and middle SNR regime, and it performs the same as BPCD by Curtailment in the high SNR regime. This shows that the Gaussian Approximation method could capture the information of the remaining Gaussian components after the curtailment in the low and middle SNR regime. 

Next, we use the observations in Fig. \ref{Fig8b} to explain the results in Fig. \ref{Fig8a}. Specifically, in the low SNR regime, the detection error is dominated by noise, and thus the detector performance is not sensitive to the estimation error; on the other hand, in the middle and high SNR regime, the detector performance is sensitive to the estimation error, and thus BPCD by Curtailment outperforms BPCD by Gaussian Approximation. In addition, BPCD Hybrid performs the best in the middle SNR regime. Nevertheless, the complexity of BPCD by Gaussian Approximation is smaller than the other two methods by \emph{GMR} times. 

\section{Detector Design without Prior Distributions of Channel Parameters}
The detector in Section IV requires knowledge of the \emph{a priori} distributions of the channels and CFOs. In this section, we extend the framework to one that does not require such prior information. The new framework is more versatile and more robust in that it can work under different possible channel and CFO distributions. We will further show that the new detector has nearly the same BER performance as the previous detector without the benefit of the knowledge of the prior distributions. Thus, the versatility and robustness of the new framework can be obtained without trading off the performance.

In the new framework, we still need a rough range of the possible CFOs. The previous framework in Section IV is still largely applicable. In the new framework, we simply remove the prior information $p({{\bf{h}}_n},\textbf{f})$  within the integrand in \eqref{eq:IV-2}. We assume the Curtailment method in all BPCDs in this section. 
\begin{figure}[ht]
	\centering
	\subfigure[BER of BPCD under Rayleigh fading channels with $\mathbb{E}\left( {{{\left| {{h_A}} \right|}^2}} \right) =\mathbb{E}\left( {{{\left| {{h_B}} \right|}^2}} \right) = 1$.]
	{
		\label{Fig9a}
		\includegraphics[width=0.47\columnwidth]{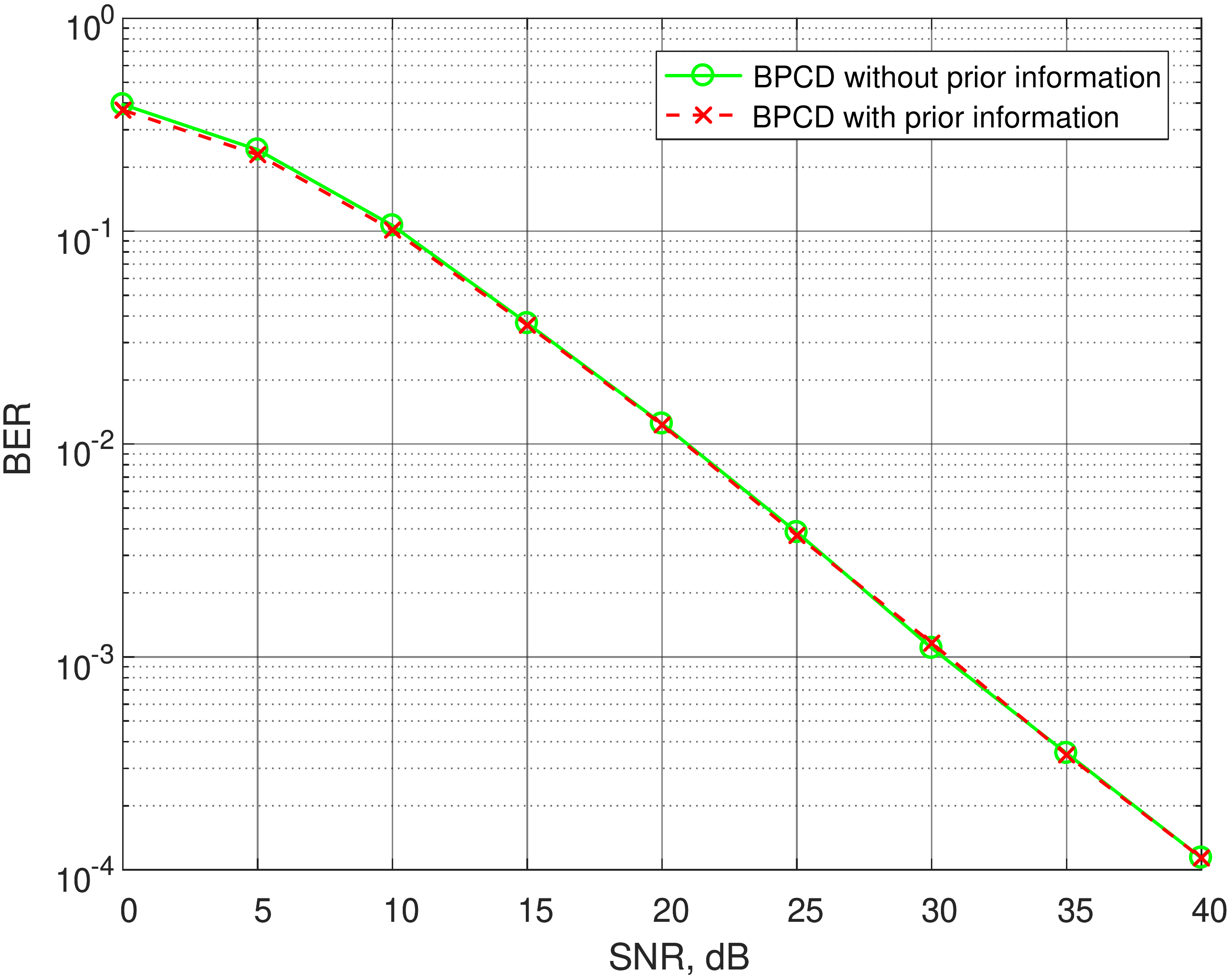}
	}
	\subfigure[BER of BPCD under channels of ${h_A} = 1$ and ${h_A} = 10$.]
	{
		\label{Fig9b}
		\includegraphics[width=0.47\columnwidth]{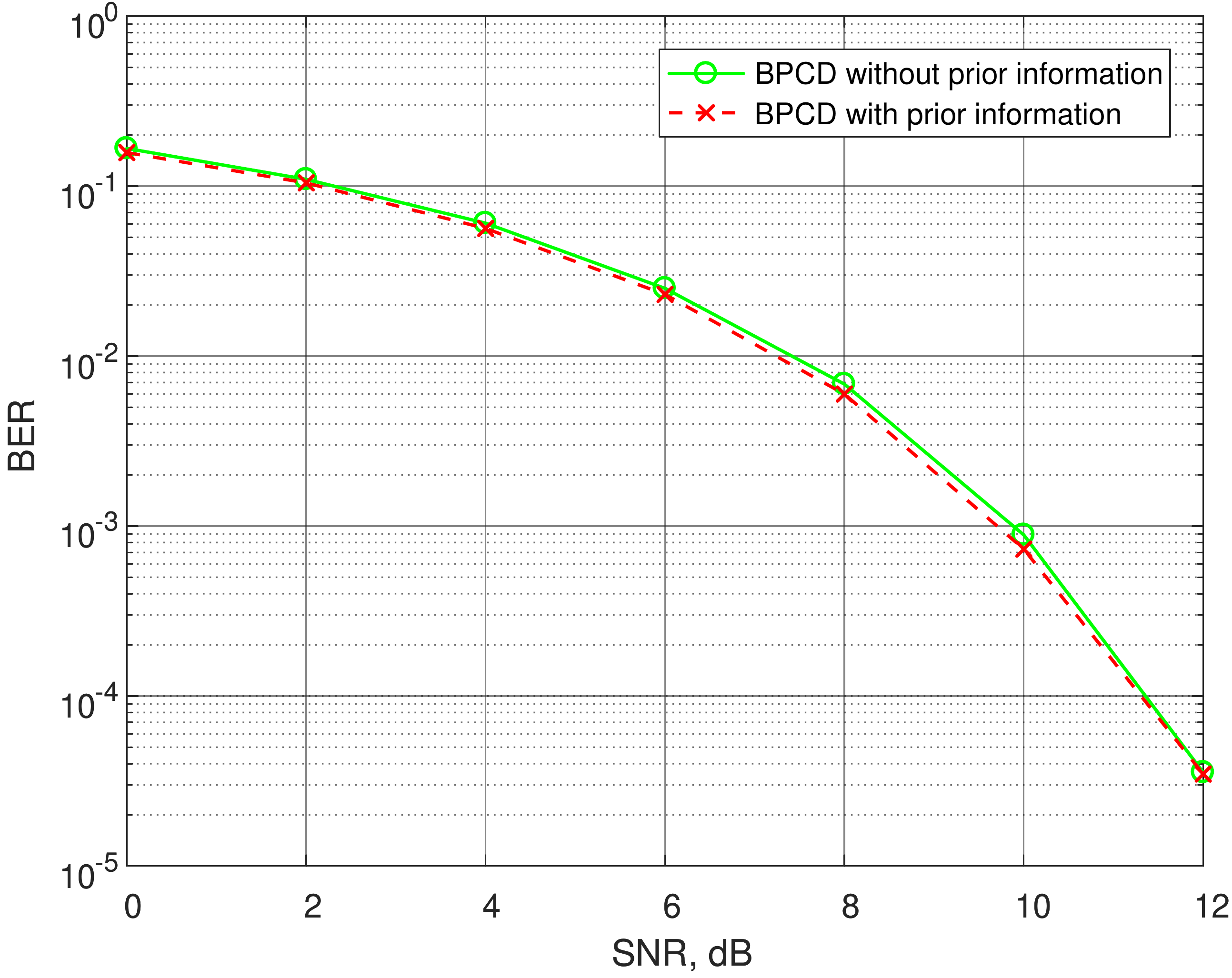}
	}
	\subfigure[BER of BPCD under channels of $p({h_A} = 1,{h_B} = 2) = 0.01$, $p({h_A} = 1,{h_B} = 3) = 0.09$,  $p({h_A} = 2,{h_B} = 2) = 0.09$,  $p({h_A} = 2,{h_B} = 3) = 0.81$.]
	{
		\label{Fig9c}
		\includegraphics[width=0.55\columnwidth]{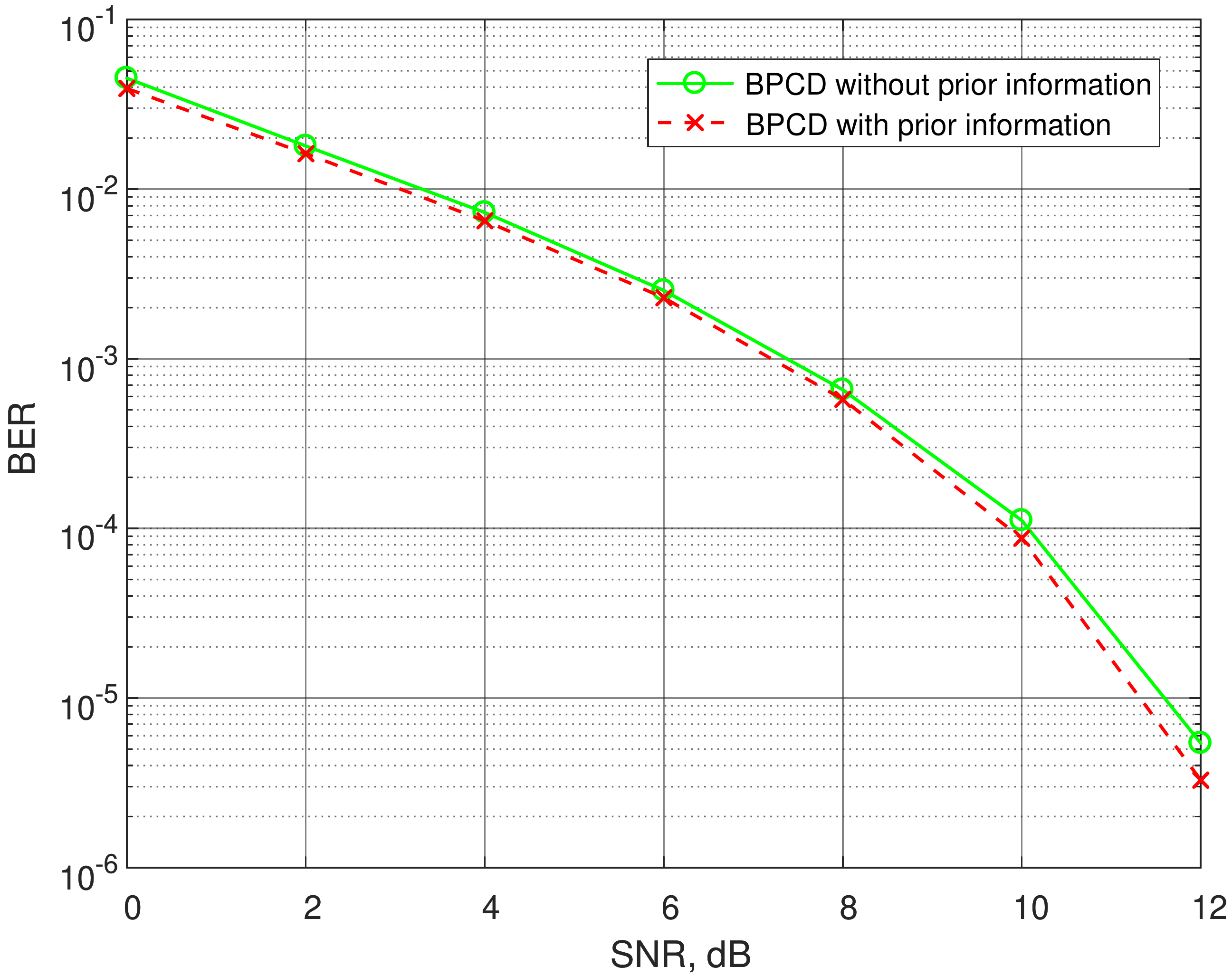}
	}
	\caption{BER of BPCD with and without the prior information. We reduce the number of Gaussian components to \emph{GMR}=4 by Curtailment. In the three cases, the distribution of CFOs is a Dirac delta function of $f_A^{{\rm{RF}}}{\rm{ = 6000}}{\kern 1pt} {\kern 1pt} {\rm{Hz}}$ and $f_B^{{\rm{RF}}}{\rm{ = 100}}{\kern 1pt} {\kern 1pt} {\rm{Hz}}$. }\label{Fig9}
\end{figure}

In the following, we present the simulation results of the new BPCD. We focus on
\begin{enumerate}
	\item comparison of the performance of the new BPCD and the previous BPCD;
	\item comparison of the performance of the new BPCD and the noncoherent detector in\cite{wang2018noncoherent} (both do not have knowledge of \emph{a priori} channel and CFO distributions) under the same set-up. In particular, we assume in the set-up that the CFO could fully characterize the symbol-to-symbol relative phase rotation between the local oscillators of user \emph{u} and relay \emph{R}. 
\end{enumerate}
For 1), our results indicate that \emph{a priori} channel distribution is not important for good detection performance. For 2), our results suggest that the coherent detector performs better than the noncoherent detector under the above set-up.
\begin{figure}[ht]
	\centering
	\subfigure[SNR=0 dB.]
	{
		\label{Fig11a}
		\includegraphics[width=0.47\columnwidth]{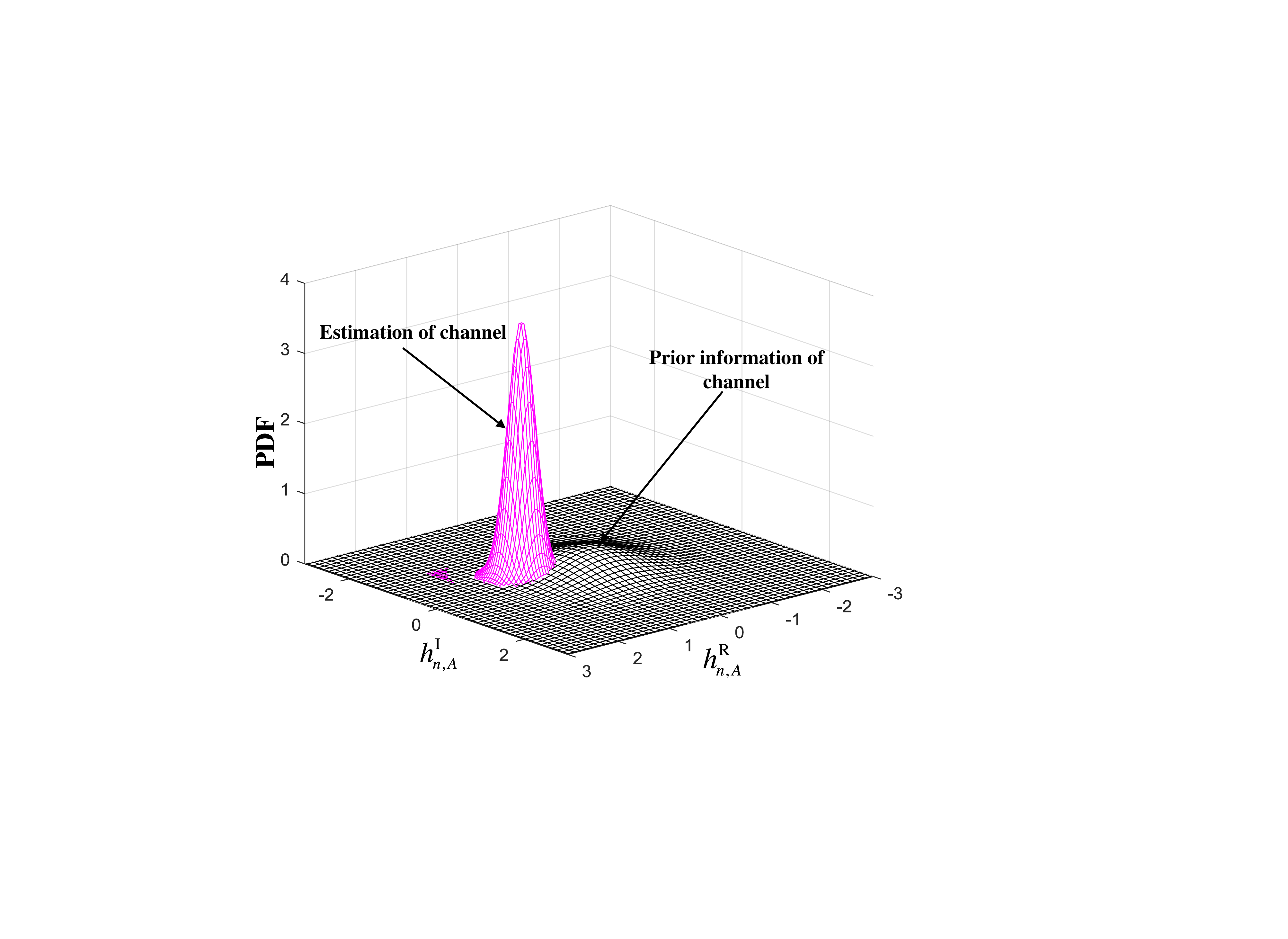}
	}
	\subfigure[SNR=10 dB.]
	{
		\label{Fig11b}
		\includegraphics[width=0.47\columnwidth]{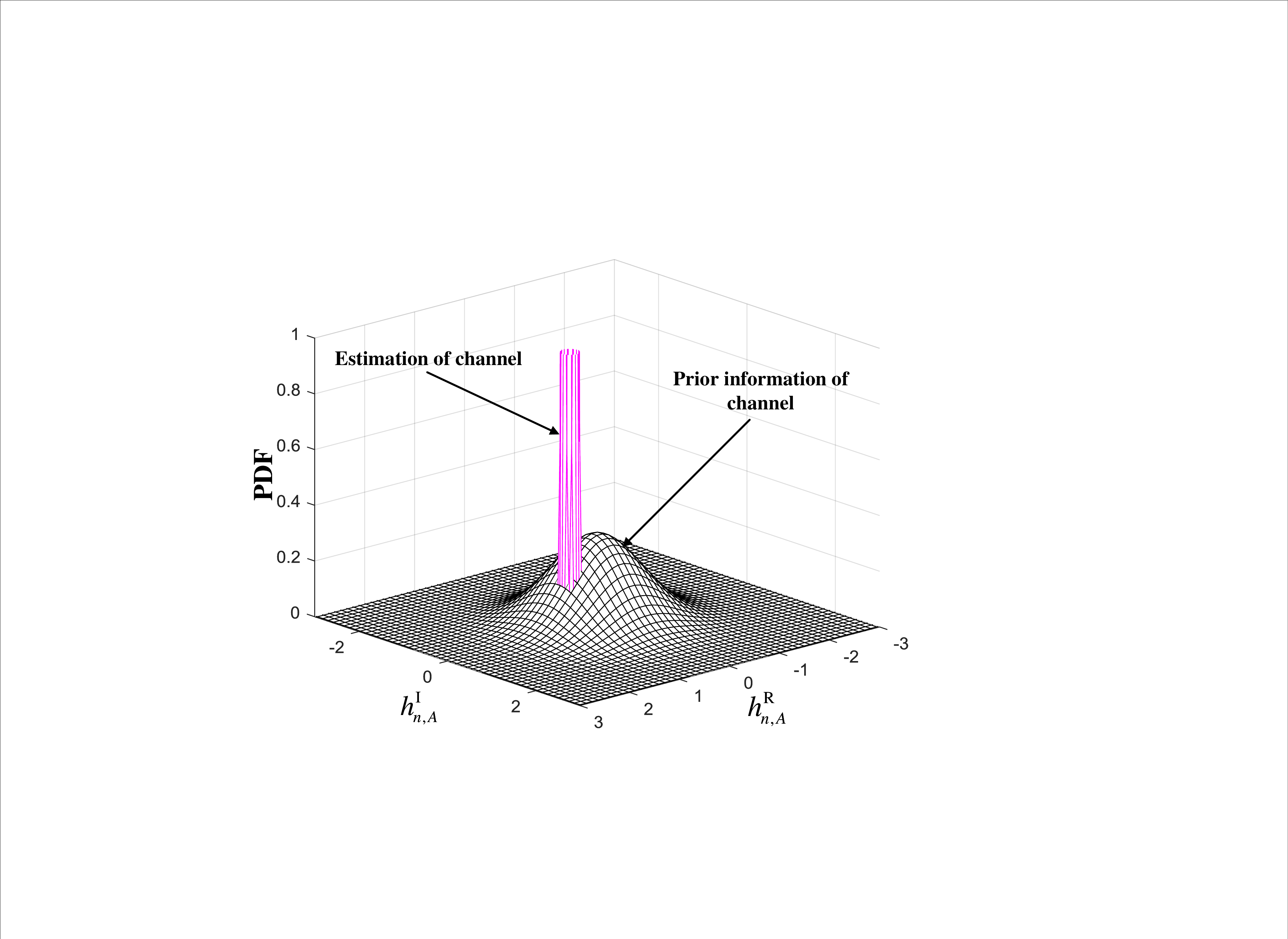}
	}
	\subfigure[SNR=40 dB.]
	{
		\label{Fig11c}
		\includegraphics[width=0.55\columnwidth]{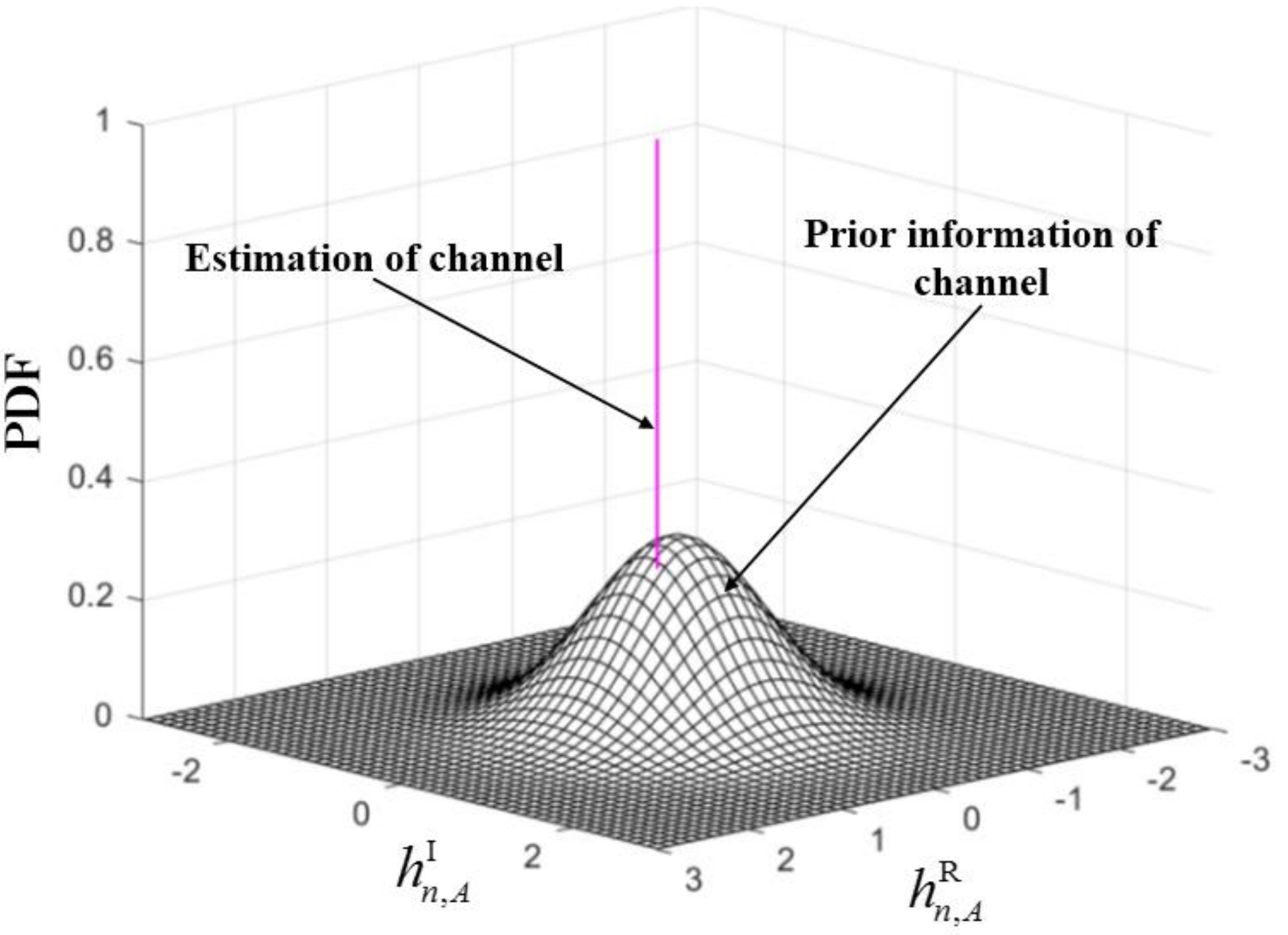}
	}
	\caption{PDF of the channel ${h_{n,A}}$  from the prior information and that from the estimation through the left-bound and right-bound messages.}\label{Fig11}
\end{figure}
 
\textbf{\emph{1) BPCD without prior information versus BPCD with prior information.}}

We first compare the performance of BPCD with and without the prior information under different distributions of the channel and CFO. For BPCD with the prior information, we feed the different prior distributions to BPCD. The results in Fig. \ref{Fig9} indicate that the two BPCD perform almost equally well. Let us explain the results with the assistance of Fig. \ref{Fig11}. We focus on a particular channel ${h_{n,A}}$  of symbol \emph{n}. As shown in Fig. \ref{Fig11}, compared with the \emph{a priori} PDF of  ${h_{n,A}}$, the PDF of  ${h_{n,A}}$ as estimated by BP is much sharper. That is, the received data symbols provide much additional and detailed information on ${h_{n,A}}$. Since the left-bound messages and right-bound messages together yield much information on the channel, prior information is not important anymore.

\textbf{\emph{2) BPCD versus noncoherent detector.}}
\begin{figure}[ht]
	\centering
	\includegraphics[scale=0.90]{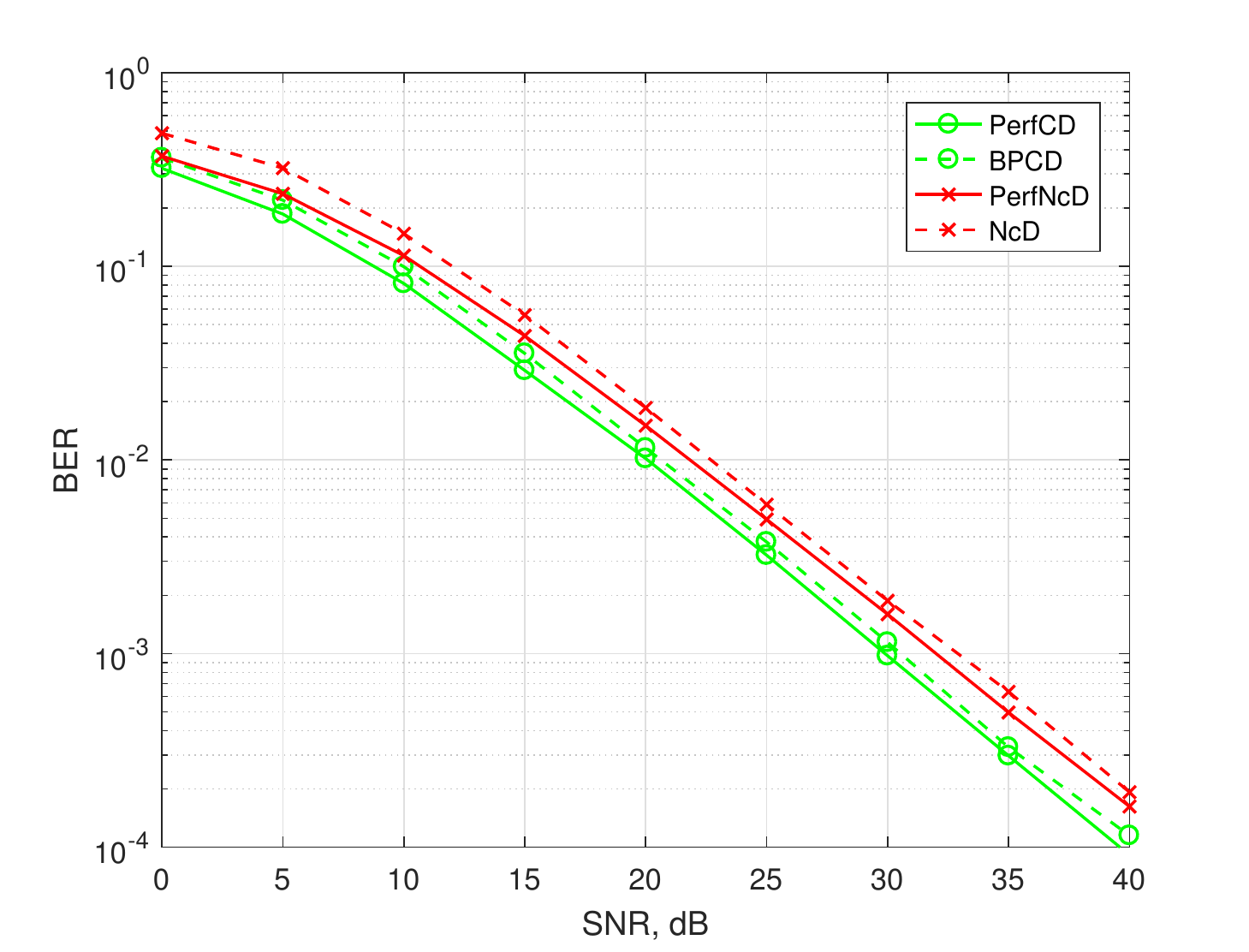}\\
	\caption{BER of PerfCD, BPCD by Curtailment with \emph{GMR}=4, PerfNcD, and NcD. The channels are Rayleigh distributed and the distribution of the CFOs is a Dirac delta function of $f_A^{{\rm{RF}}}{\rm{ = 3000}}{\kern 1pt} {\rm{Hz}}$  and $f_B^{{\rm{RF}}}{\rm{ = 1000}}{\kern 1pt} {\rm{Hz}}$.}\label{Fig12}
\end{figure}

Second, we compare the BER of BPCD in this paper and the noncoherent detector in \cite{wang2018noncoherent}. Both detectors do not have prior information on the channels and CFOs. In general, the application of the coherent detector and the noncoherent detector depends on whether only magnitude or magnitude-plus-phase of each received symbol is available in the hardware. For the former, a noncoherent detector that makes use of the magnitude information is needed; for the latter, a coherent detector that attempts to make use of the magnitude-plus-information can potentially yield better performance. We test the two detectors under the same set-up. We assume that the CFO could fully characterize the symbol-to-symbol relative phase change between the local oscillators of user $u \in \left\{ {A,B} \right\}$ and relay \emph{R}.

The noncoherent detector \cite{wang2018noncoherent} needs to estimate the \emph{relative} initial phase, \emph{relative} CFO between users \emph{A} and \emph{B}, and the magnitudes of the channel gains (i.e., $\left| {{h_A}} \right|$ and $\left| {{h_B}} \right|$) based on the magnitudes of the received signals. We refer to this detector as \textbf{``noncoherent detector (NcD)''}. In addition, we refer to the noncoherent detector with perfect knowledge of the relative initial phase, relative CFO, and the magnitudes of the channel gains as \textbf{``Perfect noncoherent detector (PerfNcD)''}.

The comparison results are shown in Fig. \ref{Fig12}, assuming the channels are Rayleigh distributed and the distribution of the CFOs is a Dirac delta function of $f_A^{{\rm{RF}}}{\rm{ = 3000}} {\kern 1pt} {\rm{Hz}}$  and $f_B^{{\rm{RF}}}{\rm{ = 1000}}{\kern 1pt} {\rm{Hz}}$. From Fig. \ref{Fig12}, the performance gap between PerfCD and PerfNcD is 2.1 dB, and the performance gap between BPCD and NcD is 2.4 dB at ${\rm{BER = 1}}{{\rm{0}}^{{\rm{ - 3}}}}$. The results suggest that the coherent detector performs better than the noncoherent detector under the above set-up. 

\section{Conclusion}
This paper investigated coherent detection for FSK-PNC with short packets. We designed a coherent detector that makes use of the BP algorithm for joint channel estimation and network-coded message detection. We showed how the BP algorithm can be simplified and made practical with Gaussian-mixture message passing. To further reduce complexity, we studied three different methods to cut down the number of Gaussian components in the mixture. 

This paper first studied a detector with knowledge on the \emph{a priori} channel distribution. Under Rayleigh fading channels, benchmarked against an ideal detector that knows the channels exactly, our detector only suffers 0.7 dB BER performance loss. This paper further studied a detector without the \emph{a priori} channel distribution. This detector is more versatile in that it can work with any channel distribution. This detector has nearly the same BER performance as the detector with prior channel distribution knowledge. In other words, the versatility of the new design can be obtained without trading off performance.

\appendices
\section{Notations}
The notations in this paper are summarized below:
\begin{basedescript}{\desclabelstyle{\pushlabel}\desclabelwidth{3.5em}}
\item[$f_u^{\rm{RF}}$:] The CFO of the RFs between user $u \in \left\{ {A,B} \right\}$  and \emph{R}.
\item[$\varphi _u^{{\rm{RF}}}$:] The initial phase offset (the phase offset at the beginning of a packet) between user \emph{u} and \emph{R}.
\item[$\varphi _{n,u}^{\rm{CFSK}}$:] The user \emph{u}'s phase accumulated over the past \emph{n}  symbol periods, assuming the use of continuous-phase FSK modulation at the transmitter. Specifically, $\varphi _{0,u}^{\rm{CFSK}} = 0$; and $\varphi _{n,u}^{\rm{CFSK}} = 2\pi \Delta fT\sum\limits_{i = 0}^{n - 1} {(2{s_{i,u}} - 1)} $ for $n \ge 1$.
\item[${\theta _{n,u}}$:] The phase offset between user $u \in \left\{ {A,B} \right\}$ and \emph{R}. Specifically, ${\theta _{n,u}} = \varphi _{n,u}^{{\rm{CFSK}}}{\rm{ + }}2n\pi f_u^{{\rm{RF}}}T + \varphi _u^{{\rm{RF}}}$.
\item[${h_u}$:] The channel between user  $u \in \left\{ {A,B} \right\}$  and \emph{R}.
\item[${h_{n,u}}$:] The resulting channel after incorporating the phases offset ${\theta _{n,u}}$ to ${h_u}$. Specifically, ${h_{n,u}} = {h_u}{e^{j{\theta _{n,u}}}}$. 
\item[${{\bf{r}}_n}$:] The \emph{n}-th received signals.
\item[${{\bf{r}}_{i:j}}$:] The vector $\left( {{{\bf{r}}_i},{{\bf{r}}_{i + 1}}, \cdots ,{{\bf{r}}_{j - 1}},{{\bf{r}}_j}} \right)$.
\end{basedescript}

\section{Validity of \eqref{eq:IV-2}}
This Appendix shows the validity of \eqref{eq:IV-2}.
Applying Bayes’ theorem on the integrand in \eqref{eq:IV-1}, we have
\begin{align}
\begin{split}
&p\left( {\left. {{{\bf{s}}_n},{{\bf{h}}_n},{\bf{f}}} \right|{{\bf{r}}_{0:N - 1}}} \right)\\
&= \sum\limits_{\scriptstyle{{\bf{s}}_{0:n - 1}}\hfill\atop
	\scriptstyle{{\bf{s}}_{n + 1,N - 1}}\hfill} {p\left( {\left. {{{\bf{s}}_{0:n - 1}},{{\bf{s}}_n},{{\bf{s}}_{n + 1,N - 1}},{{\bf{h}}_n},{\bf{f}}} \right|{{\bf{r}}_{0:N - 1}}} \right)} \\
&= \frac{1}{{p({{\bf{r}}_{0:N - 1}})}}\sum\limits_{\scriptstyle{{\bf{s}}_{0:n - 1}}\hfill\atop
	\scriptstyle{{\bf{s}}_{n + 1,N - 1}}\hfill} {p({{\bf{r}}_{0:N - 1}}|{{\bf{s}}_{0:N - 1}},{{\bf{h}}_n},{\bf{f}})p({{\bf{s}}_{0:N - 1}},{{\bf{h}}_n},{\bf{f}})} \\
&= \frac{{p({{\bf{r}}_n}|{{\bf{s}}_n},{{\bf{h}}_n})}}{{p({{\bf{r}}_{0:N - 1}})}}\sum\limits_{\scriptstyle{{\bf{s}}_{0:n - 1}}\hfill\atop
	\scriptstyle{{\bf{s}}_{n + 1,N - 1}}\hfill} {p({{\bf{r}}_{0:n - 1}},{{\bf{r}}_{n + 1:N - 1}}|{{\bf{s}}_{0:N - 1}},{{\bf{h}}_n},{\bf{f}})p({{\bf{s}}_{0:N - 1}},{{\bf{h}}_n},\bf{f})} \\
&\propto p({{\bf{h}}_n},{\bf{f}})p({{\bf{r}}_n}|{{\bf{s}}_n},{{\bf{h}}_n})\underbrace {\sum\limits_{\scriptstyle{{\bf{s}}_{0:n - 1}}\hfill\atop
		\scriptstyle{{\bf{s}}_{n + 1,N - 1}}\hfill} {p({{\bf{r}}_{0:n - 1}},{{\bf{r}}_{n + 1:N - 1}}|{{\bf{s}}_{0:N - 1}},{{\bf{h}}_n},\bf{f})} }_\Omega 
\end{split}. \label{eq:Apd-B2}
\end{align}
In the above, the fourth line follows from the third line because given  ${{\bf{s}}_n} = \left( {{s_{n,A}},{s_{n,B}}} \right)$ and ${{\bf{h}}_n}$, the term ${{\bf{Z}}_{\left( {{s_{n,A}},{s_{n,B}}} \right)}}{{\bf{h}}_n}$  in \eqref{eq:Sys-B4.1} is fixed. Thus the remaining uncertainty in ${{\bf{r}}_n}$  is the i.i.d. noise ${{\bf{w}}_n} = {{\bf{r}}_n} - {{\bf{Z}}_{\left( {{s_{n,A}},{s_{n,B}}} \right)}}{{\bf{h}}_n}$, which does not depend on ${\bf{f}}$, ${{\bf{r}}_{0:n - 1}}$, and ${{\bf{r}}_{n + 1:N - 1}}$. 

To further decompose the term $\Omega $  in \eqref{eq:Apd-B2}, we note that if ${{\bf{s}}_{n - 1}}$,  ${{\bf{h}}_n}$, and ${\bf{f}}$  are given, then ${{\bf{h}}_{n - 1}}$  can be derived from ${{\bf{h}}_n}$  exactly as in \eqref{eq:Sys-D7}. Specifically,
\begin{align}
{{\bf{h}}_{n - 1}} = {\bf{G}}_{n - 1}^{ - 1}{{\bf{h}}_n} \label{eq:Apd-B2.1}
\end{align}
where ${\bf{G}}_{n - 1}^{ - 1}$  depends on  ${{\bf{s}}_{n - 1}}$ and $\bf{f}$. Further, if ${{\bf{s}}_n}$, ${{\bf{h}}_n}$, and $\bf{f}$ are given, then we also have
\begin{align}
{{\bf{h}}_{n + 1}} = {\bf{G}}_n^{}{{\bf{h}}_n} \label{eq:Apd-B2.2}
\end{align}
where  ${\bf{G}}_n^{}$ depends on ${{\bf{s}}_n}$  and $\bf{f}$. Similarly, if ${{\bf{s}}_{0:N - 1}}$, ${{\bf{h}}_n}$, and $\bf{f}$  are given, we could have ${{\bf{h}}_0}, \cdots ,{{\bf{h}}_{n - 1}},{{\bf{h}}_{n + 1}}, \cdots ,{{\bf{h}}_{N - 1}}$  from \eqref{eq:Sys-D7}. In this case,
\begin{align}
\begin{split}
&\Omega \\
&= \sum\limits_{\scriptstyle{{\bf{s}}_{0:n - 1}}\hfill\atop
	\scriptstyle{{\bf{s}}_{n + 1,N - 1}}\hfill} {p\left( {{{\bf{r}}_{0:n - 1}},{{\bf{r}}_{n + 1:N - 1}}\left| \begin{array}{l}
		{{\bf{s}}_{0:N - 1}},{\kern 1pt} {\kern 1pt} {\kern 1pt} {{\bf{h}}_n},{\bf{f}},{{\bf{h}}_0} = {\bf{G}}_0^{ - 1}{{\bf{h}}_1}, \cdots ,{{\bf{h}}_{n - 1}} = {\bf{G}}_{n - 1}^{ - 1}{{\bf{h}}_n},\\
		{{\bf{h}}_{n + 1}} = {\bf{G}}_n^{}{{\bf{h}}_n}, \cdots , {{\bf{h}}_{N - 1}} = {\bf{G}}_{N - 2}^{}{{\bf{h}}_{N - 2}}
		\end{array} \right.} \right)} \\
&= \sum\limits_{\scriptstyle{{\bf{s}}_{0:n - 1}}\hfill\atop
	\scriptstyle{{\bf{s}}_{n + 1,N - 1}}\hfill} \begin{array}{l}
p({{\bf{r}}_{0:n - 1}},{{\bf{r}}_{n + 1:N - 1}}|{{\bf{s}}_{0:N - 1}},{\kern 1pt} {\kern 1pt} {\kern 1pt} {{\bf{h}}_{0:N - 1}},{\bf{f}})\\
\times {\psi _{{{\bf{G}}_0}}}({{\bf{h}}_1},{{\bf{h}}_0}) \cdots {\psi _{{{\bf{G}}_{n - 1}}}}({{\bf{h}}_n},{{\bf{h}}_{n - 1}}){\psi _{{{\bf{G}}_n}}}({{\bf{h}}_{n + 1}},{{\bf{h}}_n}) \cdots {\psi _{{{\bf{G}}_{N - 2}}}}({{\bf{h}}_{N - 1}},{{\bf{h}}_{N - 2}})
\end{array} \\
&= \sum\limits_{\scriptstyle{{\bf{s}}_{0:n - 1}}\hfill\atop
	\scriptstyle{{\bf{s}}_{n + 1,N - 1}}\hfill} \begin{array}{l}
p({{\bf{r}}_0}|{{\bf{h}}_0},{{\bf{s}}_0}) \cdots p({{\bf{r}}_{n - 1}}|{{\bf{h}}_{n - 1}},{{\bf{s}}_{n - 1}})p({{\bf{r}}_{n + 1}}|{{\bf{h}}_{n + 1}},{{\bf{s}}_{n + 1}}) \cdots p({{\bf{r}}_{N - 1}}|{{\bf{h}}_{N - 1}},{{\bf{s}}_{N - 1}})\\
\times {\psi _{{{\bf{G}}_0}}}({{\bf{h}}_1},{{\bf{h}}_0}) \cdots {\psi _{{{\bf{G}}_{n - 1}}}}({{\bf{h}}_n},{{\bf{h}}_{n - 1}}){\psi _{{{\bf{G}}_n}}}({{\bf{h}}_{n + 1}},{{\bf{h}}_n}) \cdots {\psi _{{{\bf{G}}_{N - 2}}}}({{\bf{h}}_{N - 1}},{{\bf{h}}_{N - 2}})
\end{array} 
\end{split} \label{eq:Apd-B3}
\end{align}
where the third line to the fourth line follows from \eqref{eq:Sys-B4.1}. In the above, ${\psi _G}(x,y)$  is an indicator function such that ${\psi _G}(x,y) = 1$  if $x = Gy$   and ${\psi _G}(x,y) = 0$  otherwise. Substituting \eqref{eq:Apd-B3} into \eqref{eq:Apd-B2}, we have
\begin{align}
\begin{split}
&p\left( {\left. {{{\bf{s}}_n},{{\bf{h}}_n},{\bf{f}}} \right|{{\bf{r}}_{0:N - 1}}} \right)\\
&\propto p({{\bf{h}}_n},{\bf{f}})\underbrace {\prod\limits_{i = {\rm{0}}}^{n - 1} {\sum\limits_{{\kern 1pt} {\kern 1pt} {\kern 1pt} {\kern 1pt} {\kern 1pt} {{\bf{s}}_i}} {p({{\bf{r}}_i}|{{\bf{h}}_i},{{\bf{s}}_i}){\psi _{{{\bf{G}}_i}}}({{\bf{h}}_{i + 1}},{{\bf{h}}_i})} } }_{{\rm M}_r^{\left( n \right)}\left( {{{\bf{h}}_n}} \right)}p({{\bf{r}}_n}|{{\bf{s}}_n},{{\bf{h}}_n})\underbrace {\prod\limits_{i = n{\rm{ + 1}}}^{N - {\rm{1}}} {\sum\limits_{{\kern 1pt} {\kern 1pt} {\kern 1pt} {\kern 1pt} {\kern 1pt} {{\bf{s}}_i}} {{\psi _{{{\bf{G}}_{i - 1}}}}({{\bf{h}}_i},{{\bf{h}}_{i - 1}})p({{\bf{r}}_i}|{{\bf{h}}_i},{{\bf{s}}_i})} } }_{{\rm M}_l^{\left( n \right)}\left( {{{\bf{s}}_n},{{\bf{h}}_n}} \right)}
\end{split} \label{eq:Apd-B4}
\end{align}
where $p({{\bf{h}}_n},\bf{f})$ is the prior information. 

\section{Expressing $p\left( {\left. {{{\bf{r}}_n}} \right|{{\bf{s}}_n},{\bf{h}}_n^{}} \right)$ as a Gaussian Function in ${\bf{h}}_n^{}$}
This appendix shows how to write $p\left( {\left. {{{\bf{r}}_n}} \right|{{\bf{s}}_n},{\bf{h}}_n^{}} \right)$  as a Gaussian function in ${\bf{h}}_n^{}$. The following details the four cases of ${{\bf{s}}_n} \in \{ (0,1),(1,0),(0,0),(1,1)\}$:
\begin{enumerate}
\item ${{\bf{s}}_n} = ({\rm{0}},{\rm{0}})$.\\
	From \eqref{eq:Sys-B4.1}, the expression of $p\left( {\left. {{{\bf{r}}_n}} \right|{{\bf{s}}_n},{\bf{h}}_n^{}} \right)$ is
	\begin{align}
	\begin{split}
	&p\left( {\left. {{{\bf{r}}_n}} \right|{{\bf{s}}_n},{\bf{h}}_n^{}} \right)\\
	&\propto \exp \left\{ { - \frac{1}{{2{N_0}}}{{\left( {\left( {\begin{array}{*{20}{c}}
						{\rm{1}}&{\rm{0}}
						\end{array}} \right){{\bf{r}}_n} - \left( {\begin{array}{*{20}{c}}
						{\rm{1}}&{\rm{1}}
						\end{array}} \right){\bf{h}}_n^{}} \right)}^ * }\left( {\left( {\begin{array}{*{20}{c}}
				{\rm{1}}&{\rm{0}}
				\end{array}} \right){{\bf{r}}_n} - \left( {\begin{array}{*{20}{c}}
				{\rm{1}}&{\rm{1}}
				\end{array}} \right){\bf{h}}_n^{}} \right)} \right\}\exp \left\{ { - \frac{1}{{2{N_0}}}{{\left| {{{\bf{r}}_{n,{\rm{2}}}}} \right|}^{\rm{2}}}} \right\}\\
	& = {c^{{{\bf{s}}_n}}}\exp \left\{ { - \frac{1}{2}{{\left( {{\bf{h}}_n^{} - {{{\bf{\bar h}}}^{{{\bf{s}}_n}}}} \right)}^ * }{{\left( {{\Sigma ^{{{\bf{s}}_n}}}} \right)}^{{\rm{ - 1}}}}\left( {{\bf{h}}_n^{} - {{{\bf{\bar h}}}^{{{\bf{s}}_n}}}} \right)} \right\}
	\end{split} \label{eq:Apd-C1}
	\end{align}
    where ${{\bf{\bar h}}^{{{\bf{s}}_n}}} = \left( {\begin{array}{*{20}{c}}
    	{\frac{{\rm{1}}}{{\rm{2}}}}&{\rm{0}}\\
    	{\frac{{\rm{1}}}{{\rm{2}}}}&{\rm{0}}
    	\end{array}} \right){{\bf{r}}_n}$,
       ${\left( {{\Sigma ^{{{\bf{s}}_n}}}} \right)^{{\rm{ - 1}}}} = \left( {\begin{array}{*{20}{c}}
    	{\frac{1}{{{N_0}}}}&{\frac{1}{{{N_0}}}}\\
    	{\frac{1}{{{N_0}}}}&{\frac{1}{{{N_0}}}}
	   \end{array}} \right)$,
   and 
   ${c^{{{\bf{s}}_n}}}{\rm{ = }}\exp \left\{ { - \frac{1}{{2{N_0}}}{{\left| {{{\bf{r}}_{n,{\rm{2}}}}} \right|}^{\rm{2}}}} \right\}$.
   Note that equations \eqref{eq:IV-C7}-\eqref{eq:IV-C9} are still valid even though ${\left( {{\Sigma ^{{{\bf{s}}_n}}}} \right)^{{\rm{ - 1}}}}$  is not invertible because the equations \eqref{eq:IV-C7}-\eqref{eq:IV-C9} does not need to invert  ${\left( {{\Sigma ^{{{\bf{s}}_n}}}} \right)^{{\rm{ - 1}}}}$.
\item ${{\bf{s}}_n} = ({\rm{1}},{\rm{1}})$.\\
   Similar to the case of ${{\bf{s}}_n} = ({\rm{0}},{\rm{0}})$, ${{\bf{\bar h}}^{{{{\bf{s}}_n}}} = \left( {\begin{array}{*{20}{c}}
   	{\rm{0}}&{\frac{{\rm{1}}}{{\rm{2}}}}\\
   	{\rm{0}}&{\frac{{\rm{1}}}{{\rm{2}}}}
   	\end{array}} \right){\bf{r}}_n}$,
   ${\left( {{\Sigma ^{{{\bf{s}}_n}}}} \right)^{{\rm{ - 1}}}} = \left( {\begin{array}{*{20}{c}}
   	{\frac{1}{{{N_0}}}}&{\frac{1}{{{N_0}}}}\\
   	{\frac{1}{{{N_0}}}}&{\frac{1}{{{N_0}}}}
   	\end{array}} \right)$,
   and
   ${c^{{{\bf{s}}_n}}}{\rm{ = }}\exp \left\{ { - \frac{1}{{2{N_0}}}{{\left| {{{\bf{r}}_{n,{\rm{1}}}}} \right|}^{\rm{2}}}} \right\}$.
\item ${{\bf{s}}_n} = ({\rm{0}},{\rm{1}})$.\\
   As a Gaussian function of ${\bf{h}}_n^{}$, $p\left( {\left. {{{\bf{r}}_n}} \right|{{\bf{s}}_n},{\bf{h}}_n^{}} \right)$ has ${{\bf{\bar h}}^{{{\bf{s}}_n}}} = {{\bf{r}}_n}$, 
   ${\left( {{\Sigma ^{{{\bf{s}}_n}}}} \right)^{{\rm{ - 1}}}} = \left( {\begin{array}{*{20}{c}}
   	{\frac{{\rm{1}}}{{{N_{\rm{0}}}}}}&0\\
   	0&{\frac{{\rm{1}}}{{{N_{\rm{0}}}}}}
   	\end{array}} \right)$,
   and ${c^{{{\bf{s}}_n}}}{\rm{ = 1}}$.
\item ${{\bf{s}}_n} = ({\rm{1}},{\rm{0}})$.\\
    As a Gaussian function of ${\bf{h}}_n^{}$, $p\left( {\left. {{{\bf{r}}_n}} \right|{{\bf{s}}_n},{\bf{h}}_n^{}} \right)$ has 
    ${{\bf{\bar h}}^{{{\bf{s}}_n}}} = \left( {\begin{array}{*{20}{c}}
    	0&1\\
    	1&0
    	\end{array}} \right){{\bf{r}}_n}$,
    ${\left( {{\Sigma ^{{{\bf{s}}_n}}}} \right)^{{\rm{ - 1}}}} = \left( {\begin{array}{*{20}{c}}
    	{\frac{{\rm{1}}}{{{N_{\rm{0}}}}}}&0\\
    	0&{\frac{{\rm{1}}}{{{N_{\rm{0}}}}}}
    	\end{array}} \right)$,
    and ${c^{{{\bf{s}}_n}}}{\rm{ = 1}}$.
\end{enumerate}

\bibliographystyle{IEEEtran}
\bibliography{references}
\end{document}